\documentclass{aa} 
\usepackage{graphicx}
\usepackage{txfonts}
\usepackage{rotating}
\usepackage{natbib}
\newcommand{\asec}{$^{\prime\prime}$}
\newcommand{\pas}{.\hskip-2pt$^{\prime\prime}$}

\def\i{I05345}
\def\H{N$_{2}$H$^{+}$}
\def\D{N$_{2}$D$^{+}$}

\def\AMM{NH$_3$}

\def\CI{\mbox{$^{13}$CO}}
\def\CII{\mbox{C$^{18}$O}}

\def\CO{\mbox{$^{12}$CO}}

\def\HII{H{\sc ii}}

\def\kms{\mbox{km~s$^{-1}$}}
\def\cmc{cm$^{-3}$}
\def\cmq{cm$^{-2}$}

\def\solm{\mbox{M$_\odot$}}

\def\Vlsr{$V_{\rm LSR}$}
\def\Dfrac{$D_{\rm frac}$}

\begin{document}

\title{Linking pre-- and proto--stellar objects in the intermediate-/high-mass star forming 
region IRAS 05345+3157
\thanks{Based on observations carried out with the IRAM Plateau de Bure Interferometer.
IRAM is supported by INSU/CNRS (France), MPG (Germany) and IGN (Spain).}
\thanks{The Submillimeter Array 
is a joint project between
the Smithsonian Astrophysical Observatory and the Academia Sinica Institute
of Astronomy and Astrophysics, and is funded by the Smithsonian Institution 
and the Academia Sinica.}}
\author{F. Fontani \inst{1,2} \and  Q. Zhang \inst{3} \and P. Caselli \inst{4} \and
	T. L. Bourke \inst{3}
        }
\offprints{F. Fontani, \email{Francesco.Fontani@unige.ch}}
\institute{ISDC, Ch. d'Ecogia 16, 1290 Versoix, Switzerland
	\and
	   Observatoire de Gen\`eve, University of Geneva, Ch. de Maillettes 51, 1290 Sauverny, Switzerland 
	   \and
	   Harvard-Smithsonian Center for Astrophysics, 60 Garden Street, Cambridge, MA 02138, USA
	   \and
	   School of Physics and Astronomy, University of Leeds, Leeds, LS2 9JT, UK
	   } 
\date{Received date; accepted date}

\titlerunning{The intermediate-/high-mass protocluster in I05345}
\authorrunning{Fontani et al.}

\abstract
{To better understand the initial conditions of the high-mass
star formation process, it is crucial to study at high-angular resolution the morphology, the
kinematics, and eventually the interactions of the coldest condensations 
associated with intermediate-/high-mass star forming regions.}
{The paper studies the cold condensations in the intermediate-/high-mass 
proto-cluster IRAS 05345+3157, focusing the attention on the interaction with the other 
objects in the cluster.}
{We have performed millimeter high-angular resolution observations, both
in the continuum and several molecular lines, with the PdBI and the SMA.
In a recent paper, we have already published part of these data. The main finding
of that work was the detection of two cold and dense gaseous condensations, 
called N and S (masses $\sim 2$ and $\sim 9$ \solm ), characterised by
high values of the deuterium fractionation ($\sim 0.1$ in both cores) obtained 
from the column density ratio $N$(\D )/$N$(\H ).  In this paper, we present 
a full report of the observations, and a complete analysis of the data obtained.}
{The millimeter maps reveal the presence of 3 cores inside the interferometers
primary beam, called C1-a, C1-b and C2. None of them are associated
with cores N and S. C1-b is very likely associated with a newly formed
early-B ZAMS star embedded inside a hot-core, while C1-a is more likely 
associated with a class 0 intermediate-mass protostar. The nature of C2 is unclear.
Both C1-a and C1-b are good candidates as driving sources of a powerful \CO\ outflow,
which strongly interacts with N, as demonstrated by the velocity gradient of the gas along 
this condensation. The \H\ linewidths are between $\sim 1$ and 2 \kms\ in the region where
the continuum cores are located, and smaller ($\sim 0.5 - 1.5$ \kms) towards
N and S, indicating that the gas in the deuterated condensations is
more quiescent than that associated with the continuum sources.
This is consistent with the fact that they are still
in the pre--stellar phase and hence the star formation process has not yet taken 
place in there.}
{The study of the gas kinematics across the source indicates a tight
interaction between the deuterated condensations and the sources 
embedded in the millimeter cores. For the nature of N and S, we propose two 
scenarios: they can be low-mass pre--stellar condensations or 'seeds' of future 
high-mass star(s). From these data however it is not possible to establish how 
the turbulence triggered by the neghbouring cluster of protostars can influence the 
evolution of the condensations.}
\keywords{Stars: formation -- Radio lines: ISM  -- ISM: individual (IRAS 05345+3157) -- ISM: molecules}

\maketitle
%

\section{Introduction}
\label{Introduction}

The initial conditions of the star formation process are still a
matter of debate. Studies have begun to unveil the chemical and physical 
properties of starless low-mass cores on the verge of forming 
low-mass stars (Kuiper et al.~\citeyear{kuiper}; Caselli 
et al.~\citeyear{casellia},~\citeyear{casellib}; 
Tafalla et al.~\citeyear{tafalla02},~\citeyear{tafalla06}),
demonstrating that in the dense and cold nuclei of these cores,
C-bearing molecular species such as CO and CS are strongly depleted
(e.g.~Caselli et al.~\citeyear{casellib}; Tafalla et al.~\citeyear{tafalla02}),
while N-bearing molecular ions such as \D\ and \H\
(Caselli et al.~2002a, 2002b; Crapsi et al.~2005) maintain
large abundances in the gas phase, and their column 
density ratios reach values of $\sim 0.1$ or more, much higher than 
the cosmic [D/H] elemental abundance ($\sim 10^{-5}$, Oliveira et al.~2003).

The characterisation of the earliest stages of the formation process of high-mass 
stars is more difficult than for low-mass objects, given their shorter evolutionary 
timescales, larger distances, and strong interaction with their 
environments. Nevertheless, significant progress has been made
by various authors who have performed extensive 
studies aimed at the identification of massive protostellar candidates, 
i.e. very young ($<10^{5}$ yr), massive (M$>8$\solm ) stellar
objects which have not yet ionised the surrounding medium
(Molinari et al.~\citeyear{mol96}; 
Sridharan et al.~\citeyear{sridharan}; Fontani et al.~\citeyear{fonta05}).
On the other hand, the pre--protostellar stage, namely the
phase in which a cold starless core evolves towards the
onset of the formation of a high-mass protostar, has not yet 
been investigated in great detail, even though in the last
few years an increasing effort has been devoted to this study (see e.g.~Wang
et al.~\citeyear{wang}; Zhang et al.~\citeyear{zhang09}; Beuther \& Sridharan~\citeyear{bes}). 

Many ultracompact (UC) \HII\ regions are located in clusters 
(see e.g. Thompson et al.~\citeyear{thompson}), and it is
common to find massive young stellar objects in
different evolutionary stages in the same star forming region (Kurtz
et al.~\citeyear{kurtz}). Therefore,
one expects to find even earlier phases of massive star formation 
in the vicinity of UC \HII\ regions and/or other ``signposts''
of massive star formation activity.
With this in mind, several authors have recently searched for cold
and dense spots in the neighbourhood of luminous IRAS point sources
(Fontani et al.~\citeyear{fonta06}), UC \HII\ regions 
(Pillai et al.~\citeyear{pillai}) and methanol masers 
(Hill et al.~\citeyear{hill}). In particular, Fontani et 
al.~(\citeyear{fonta06}) have searched for \D\ emission
in the gas associated with 10 luminous IRAS sources
with the IRAM-30m telescope, and 
detected deuterated gas in 7 out of the 10 sources observed.

The object of the present work is one of the sources
studied by Fontani et al.~(\citeyear{fonta06}), which stands out
for its very interesting characteristics: it is a high-mass 
($\sim 180$\solm , Fontani et al.~\citeyear{fonta06}) dusty
clump, undetected in the MSX 8$\mu$m band, located 
nearby the luminous IRAS point source 05345+3157 ($\sim 60$\arcsec\
to the N-E). The region is located
at a distance of 1.8 kpc (Zhang et al.~\citeyear{zhang}),
and its surface density ($\sim 1.3$ gr cm$^{-2}$) and
mass-to-luminosity ratio ($\sim 8 L_{\odot}/M_{\odot}$)
indicate that it is potentially a site of massive
star formation (see Fig.~1 of Chakrabarti \& McKee~\citeyear{chak}).
Hereafter, we will call our target \i .
Previous interferometric studies have reavealed a clumpy
structure of the molecular gas in \i\ (Molinari et al.~\citeyear{mol02}).
From IRAM-30m data, Fontani et al.~(\citeyear{fonta06}) have measured 
an average CO depletion factor (ratio between expected and 
observed CO abundance) of $\sim 3$ and an average 
deuterium fractionation (the column density ratio between
a deuterated species and the corresponding one containing hydrogen) 
of $\sim 0.01$, three orders of magnitude higher than the cosmic
[D/H] abundance. These findings would indicate for the
very first time the possible presence of a {\it high-mass pre--stellar core}
(with $T\sim 10$ K and $n_{\rm H_{2}}\sim 10^{6}$\cmc\ ), analogous
to those detected in several low-mass star forming regions.
However, the angular resolution was insufficient to determine
whether we are dealing with a single high-mass core or instead with
a sample of low-mass ones.

\begin{figure*}
\centerline{\includegraphics[angle=-90,width=10cm]{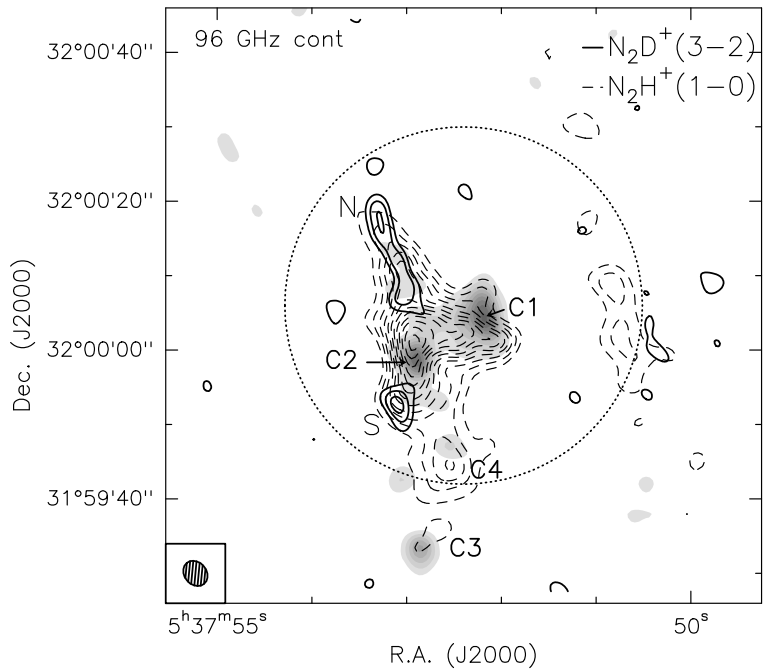}}
\caption{Summary of the main findings of paper I: the dashed
contours represent the intensity of the \H\ (1--0) line
 integrated between --21.2 and --14.5 \kms , corresponding
to the main group of the hyperfine components of this line, 
observed with the PdBI (see also Fig.~2 of paper I). Levels range from
the 3$\sigma$ rms, which is $\sim 0.02$ Jy beam$^{-1}$, to 0.26 
Jy beam$^{-1}$, in steps of 3$\sigma$. 
The grey scale shows the 96 GHz continuum: the levels range from
the 3$\sigma$ rms (4.2$\times 10^{-4}$ Jy beam$^{-1}$) to 4$\times 10^{-3}$ 
Jy beam$^{-1}$, in steps of 3$\sigma$. The solid contours represent
the \D\ (3--2) line emission integrated between --18.37 and --16.7 \kms , 
obtained with the SMA (see also Fig.~1 of paper I). The two main condensations
are indicated as N and S. Contour levels start from the 
3$\sigma$ rms ($\sim 0.09$ Jy beam$^{-1}$), and are in steps of 2 $\sigma$. The
96 GHz continuum cores are indicated as C1, C2, C3 and C4. The dotted circle 
represents the PdBI primary beam at 96 GHz ($\sim 48$\asec ). 
The ellipse in the bottom left corner shows the synthesised beam of 
the \D\ image, comparable to that of the PdBI at 96 GHz.  }
\label{fig_paperI}
\end{figure*}

Therefore, we have recently mapped \i\ at high-angular resolution in 
the \H\ (1--0) line with the IRAM Plateau de Bure Interferometer
(PdBI), and in the \H\ and  \D\ (3--2) lines with the Submillimeter Array (SMA),
in order to derive a detailed map of the deuterium fractionation in the
source.
Simultaneously, we have obtained observations in the continuum at 
$\sim 96$, $\sim 225$ and $\sim 284$ GHz with the two interferometers,
as well as in several lines of other molecules. The preliminary results of these observations, 
focused mainly on the deuterium fractionation derived from the \H\ and \D\ data, have been 
published in a recent work (Fontani et al.~\citeyear{fonta08}, hereafter paper I).
For completeness, in Fig.~\ref{fig_paperI} we summarise the main
observational results of that work, by showing: (i) the 3~mm 
continuum observed with PdBI, which reveals the presence of 4 cores,
 one of which outside the interferometer primary beam. 
(ii) the distribution of the intensity of the
\H\ (1--0) line, which is extended and with a complex structure;
(iii) the distribution of the \D\ (3--2) line integrated emission,
concentrated in two condensations, called N and
S. The integrated emission of the \H\ (3--2) line (not shown in Fig.~\ref{fig_paperI}) 
is compact, and detected towards the strongest continuum peak only.
In paper I we have derived the masses of N and S,
which are $\sim 9$ and $\sim 2.5$ \solm , respectively. Also,
from the \D\ /\H\ column density ratio we 
have obtained a deuterium fractionation of $\sim 0.1$ in both 
condensations, which are the typical values
derived in low-mass pre--stellar cores.

In this work we present a full report of the observations
presented in paper I, and a complete analysis of the data obtained.  
In Sect.~\ref{obs} we describe the 
observations and the data reduction, while the results are presented 
in Sect.~\ref{results}. In Sect.~\ref{phy_par} we derive the main
physical parameters, which are then discussed in Sect.~\ref{discu}. 
The main findings of this work are summarised in Sect.~\ref{conc}.

\section{Observations and data reduction}
\label{obs}

\begin{figure*}
\centerline{\includegraphics[angle=0,width=14cm]{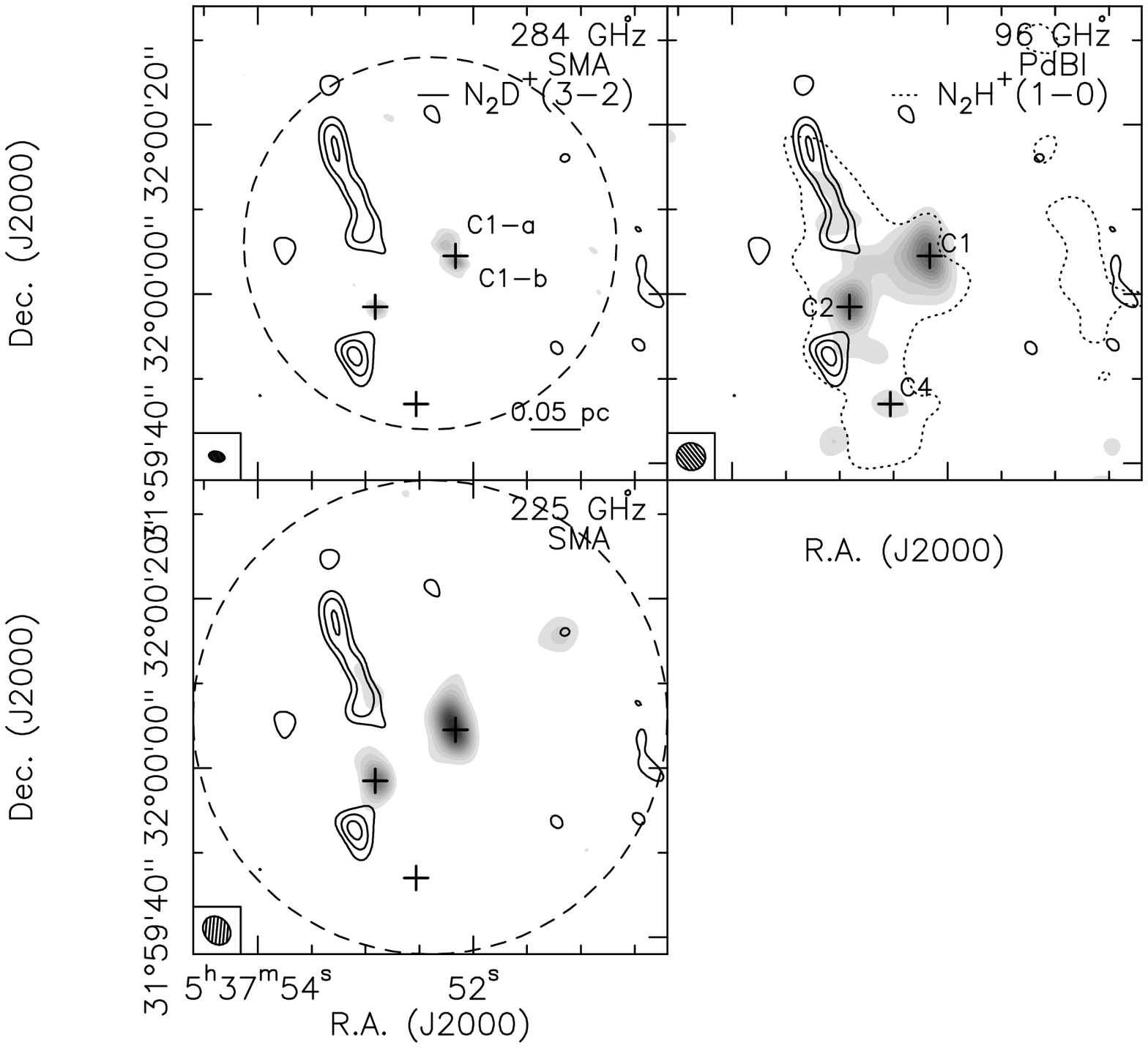}}
\caption{Top left panel: map of the 284 GHz continuum (grey scale) 
obtained with the Submillimeter Array. The first level is the 
3$\sigma$ rms = 0.018 Jy beam $^{-1}$ (step = 3$\sigma$ rms).
The crosses indicate the position of the 96 GHz continuum sources
shown in Fig.~\ref{fig_paperI}, except C3 which is outside the 
interferometers primary beam.
The thick contours correspond to the integrated emission of the \D\ (3--2) line,
starting from the 3$\sigma$ level and separated by 2$\sigma$, and thus 
indicate the position of the pre--stellar
cores detected in paper I (see Fig.~\ref{fig_paperI}). 
The ellipse in the bottom left corner represents the synthesized
beam of the continuum image. 
Bottom left panel: same as top left for the 225 GHz image. The
contours start from the 3$\sigma$ rms (= 0.006 Jy beam$^{-1}$),
and are in steps of 3$\sigma$. 
Right panel: same as left panel for the 96 GHz continuum,
observed with the PdBI. The levels are the same as in Fig.~\ref{fig_paperI}.
The dotted contour corresponds
to the 3$\sigma$ contour level of the \H\ (1--0) integrated map
shown in Fig.~1.}
\label{cont_fig}
\end{figure*}

\subsection{Submillimeter Array observations}
\label{sma}

Observations of \D\ (3--2) (at 231321 MHz) and \H\ (3--2) (at 279511.7 MHz)
towards \i\ were carried out with the SMA (Ho et al.~\citeyear{ho}) in the compact 
configuration on 30 January and 21 February 2007, respectively. 
The correlator was configured to observe
the continuum emission and several other molecular
lines simultaneously (see Table~\ref{tab_mol}). The phase centre was the nominal 
position of the sub-mm peak detected with the JCMT (Fontani et al.~2006), 
namely R.A.(J2000)=05$^h$37$^m$52.4$^s$
and Dec.(J2000)=32$^{\circ}$00$^{\prime}$06\asec , and the local
standard of rest velocity \Vlsr\ is $-18.4$ \kms .
For gain calibrations, observations of \i\ were alternated
with the quasars 3C111 and J0530+135. Quasar 3C279 and Callisto 
were used for passband and flux calibration, respectively.
The SMA data were calibrated with the MIR package (Qi~2005),
and imaged with MIRIAD (Sault et al.~\citeyear{sault}). 
Channel maps were created with natural weighting, attaining a 
resolution of: 3\pas7$\times$3\pas0 for the \D\ (3--2) 
channel map; 3\pas0$\times$2\pas8 for the 225 GHz continuum image;
2\pas7$\times$2\pas0 for the \H\ (3--2) channel map;
1\pas9$\times$1\pas2 for the 284 GHz continuum image.
At the assumed distance of 1.8~kpc, these values translate
into spatial resolutions of $\sim 0.025$ and $\sim 0.01$ pc at 
225 GHz and 284 GHz, respectively (i.e. 5000 and 2000 A.U.,
respectively).
The 3$\sigma$ level in the 225 and 284 GHz continuum images is
0.006 and 0.018 Jy beam$^{-1}$, respectively. 

The spectral resolution in velocity for the \D\ and \H\ (3--2)
was $\sim $0.53 and $\sim $0.44 \kms , respectively.
All the lines observed are listed in Table~\ref{tab_mol}:
in Cols.~1 and 2 we give the transitions observed and the line 
rest frequency, respectively; Col.~3 gives the spectral resolution;
Col.~4 reports on the detection (Y) or 
non-detection (N) of the transition, and the 3$\sigma$ level
in the channel maps of the detected transitions is given
in Col.~5. 

\begin{table*}
\begin{center}
\caption[]{Molecular transitions observed with the SMA and the PdBI.}
\label{tab_mol}
\begin{tabular}{ccccc}
\hline \hline
Transition & $\nu$ & Spec. Resol. & Detection & 3$\sigma^{\bf a}$ \\ 
           &  (GHz) &   (\kms )  &   &  (Jy beam$^{-1}$) \\
           \hline
       \multicolumn{5}{c}{ SMA } \\
\CII\ (2--1) & 219.560 & 0.55 & Y &  0.18 \\
 HNCO (10--9)  & 219.798 & 0.55 & N &  \\
 H$_2$C$^{13}$O (3$_{1,2}$--2$_{1,1}$) & 219.909 & 0.55 & N &  \\ 
SO (5$_6$--4$_5$) & 219.949 & 0.55 & Y &  0.12 \\
CH$_3$OH (8$_{0,3}$--7$_{1,3}$) & 220.079 & 0.55 & Y & 0.15 \\
HCOOCH$_3$ (17--16) & 220.167 & 0.55 & N &  \\
\CI\ (2--1) & 220.399 & 0.55 & Y &  0.16  \\
CH$_3$CN (12--11) & 220.539 -- 220.747$^{b}$ & 0.55 & Y &  0.15 \\
HC$_5$N (83--82) & 220.937 & 0.55 & N &  \\
CH$_3$OH (8$_{-1,8}$--7$_{0,7}$) & 229.759 & 0.53 & Y &  0.18 \\
CH$_3$OH (3$_{2,4}$--4$_{1,4}$) & 230.027 & 0.53 & N & \\
\CO\ (2--1) & 230.538 & 1.0 & Y & 0.15 \\
OCS (19--18) & 231.062 & 0.50 & Y &  0.21 \\
$^{13}$CS (5--4) & 231.220 & 0.50 & Y &  0.21 \\
CH$_3$OH (10$_{2,2}$--9$_{3,2}$) & 231.286 & 0.50 & N & \\
\D\ (3--2) & 231.321 & 0.52 & Y &  0.21 \\
CH$_3$OH (9$_{1,9}$--8$_{0,8}$) & 278.305 & 0.44 & N & \\
\H\ (3--2) & 279.512 & 0.44 & Y &  1.0 \\
OCS (23--22) & 279.685 & 0.44 & N & \\
DCO$^+$ (4--3) & 288.144 & 0.44 & N & \\
DCN (4--3) & 288.664 & 0.44 & N & \\
C$^{34}$S (6--5) & 289.209 & 0.44 & N & \\
CH$_3$OH (6--5) & 289.307 -- 290.939$^{c}$ & 0.44 & N &  \\
SiS (16--15) & 290.380 & 0.44 & N &  \\
\hline
\multicolumn{5}{c}{PdBI}  \\
\H\ (1--0) & 93.173 & 0.25 & Y &  0.03 \\
\hline
\end{tabular}
\end{center}
$^a$: for the detected transitions only \\
 $^b$: minimum and maximum frequencies corresponding to the components K=7 and K=0, respectively \\
 $^c$: minimum and maximum frequencies corresponding to the components ($6_{0,6}-5_{0,5}$) and ($6_{+2,4}-5_{+2,3}$), respectively
\end{table*}

\subsection{Plateau de Bure Interferometer observations}
\label{pdb}

We observed the \H\ (1--0) line at 93173.7725 MHz towards \i\ with the 
PdBI on 11 and 20 August 2006, in the D 
configuration, and on 3 April 2007 in the C configuration.
In Table~\ref{tab_mol}, the spectral resolution in velocity is given.
We used the same phase reference and \Vlsr\ velocities as for
the SMA observations. The nearby point sources 0507+179 and 0552+398 were 
used as phase calibrators, while bandpass and flux scale were 
calibrated from observations of 3C345 and MWC349,
respectively. For continuum measurements, we placed two 
320 MHz correlator units in the band when making the observations in
D configuration, and six 320 MHz correlator units in C configuration. 
The \H\ lines were excluded in averaging these units to produce the 
final continuum image (at $\sim 96095$ MHz). The synthesised beam size of
the \H\ channel map was 3\pas2$\times$3\pas4, while that of the 
continuum was 3\pas1$\times$3\pas2 , which translates into a
spatial resolution of $\sim 0.03$ pc. 
The 3$\sigma$ level in the continuum image is $4.2\times 10^{-4}$
Jy beam$^{-1}$.
We stress that the observations of the \H\ (1--0) and \D\ (3--2) lines have
approximately the same angular resolution.
The data have been reduced with the GILDAS software,
developed at IRAM and the Observatoire de Grenoble.

\subsection{Combining SMA and NRAO data for \CO\ (2--1)}
\label{datacomb}

The problem of the missing short spacing information on the SMA
data of the \CO\ (2--1) line can be solved combining the SMA data
with the peviously published \CO\ (2--1) map from the NRAO-12m telescope (Zhang et
al.~\citeyear{zhang}). For observing details on the single-dish observations
we refer to the paper by Zhang et al.~(\citeyear{zhang}). The single-dish
data were converted to visibilities in the MIRIAD package with the
task UVMODEL. The interferometer and single-dish data were then
processed together. The synthesised beam of the combined data
is  2.87\asec$\times$3.60\asec\ (PA $\sim $ 42\degr ).
For more details on the method adopted
see e.g.~Zhang et al.~(\citeyear{zhang07}).

\subsection{Data reduction}
\label{datared}

The \H\ (1--0), (3--2), and \D\ (3--2) rotational transitions 
have hyperfine structures. To take this into account, we fitted the 
lines using METHOD HFS of the CLASS program, which is part of the GAG software
developed at IRAM and the Observatoire de Grenoble. This method assumes
that all the hyperfine components have the same excitation temperature
and width, and that their separation is fixed to the laboratory value.
The method also provides an estimate of the total optical depth of
the lines, based on the intensity ratio of the different hyperfine
components. For the rest frequencies of the \H\ and \D\ lines, we have used 
the laboratory values given in Crapsi et al.~(\citeyear{crapsi}). We have not
applied the corrections given in Pagani et al.~(\citeyear{pagani}), since
these are well below the spectral resolution of our observations.

\begin{figure*}
\centerline{\includegraphics[angle=-90,width=14cm]{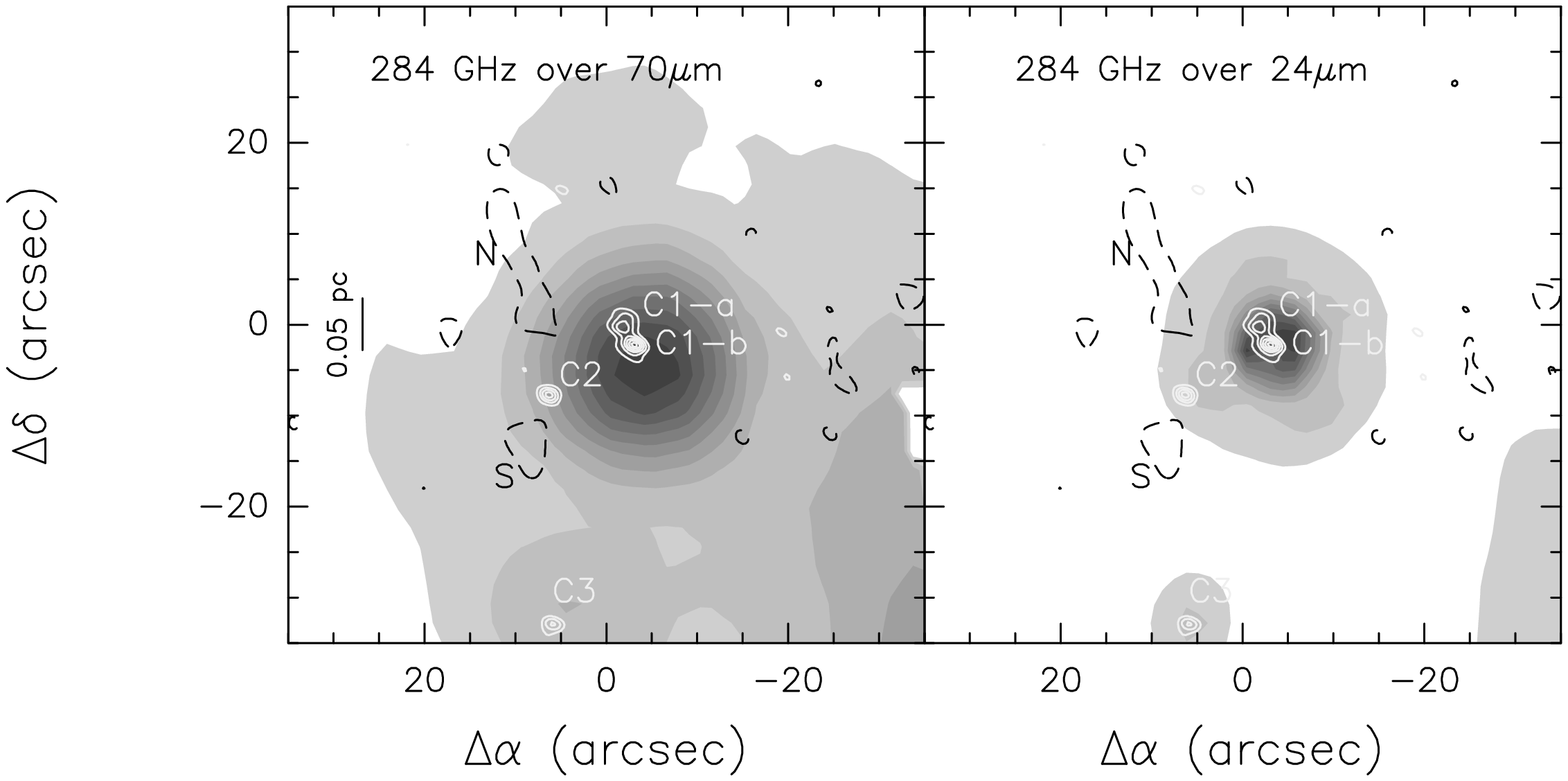}}
\caption{Left panel: map of the 284 GHz continuum (solid white contours) 
obtained with the Submillimeter Array,
superimposed on the Spitzer MIPS image of \i\ at 70 $\mu$m,
shown by the grey scale (first level and step = 10$\%$ of
the maximum). The levels of the 284 GHz image are the same
as in Fig.~\ref{cont_fig}.
The millimeter continuum sources C1-a, C1-b and C2 are labelled as
in Fig.~\ref{cont_fig}. The dashed contour represents the 3$\sigma$
level of the \D\ (3--2) line integrated emission (see Fig.~\ref{cont_fig}).
Right panel: same as left panel for the 24 $\mu$m Spitzer MIPS image.}
\label{cont_spi_fig}
\end{figure*}

\section{Results}
\label{results}

\subsection{Continuum maps}
\label{cont}
 
The 284 and 225 GHz continuum images obtained with the SMA are shown 
in top and bottom left panels of Fig.~\ref{cont_fig}, respectively.
For a complete comparison among the three continuum maps, 
 in Fig.~\ref{cont_fig} we also show the continuum emission at 96 GHz, 
already presented in Fig.~\ref{fig_paperI}.

At 284 GHz, three main compact cores are detected inside the
SMA primary beam (44\asec ), two of them close to the 
map centre and the other one is $\sim 10$\asec\ S--E of the map centre.
We will call the central condensations C1-a and C1-b, respectively,
because they are located at the position of source C1 in the 96 GHz image, 
while the eastern one exactly matches the emission peak of core C2. Among the other 96 GHz 
continuum sources shown in Fig~\ref{fig_paperI}, we clearly detect C3 outside the 
interferometer primary beam (not shown in Fig.~\ref{cont_fig}), while C4 is undetected.

At 225 GHz, the continuum sources C1-a and C1-b are blended
into C1 as in the 96 GHz image (Fig.~\ref{fig_paperI}), and C2 is clearly detected
as well. As for the 284 GHz image, C4 is 
undetected while C3 is detected but not shown because outside the
primary beam, corresponding to the interferometer field
of view.
There is a faint emission $\sim 18$\asec\ to the NE of C1-a and
C1-b, but we decided not to consider this as a real source because it
is detected at 6$\sigma$ in this image only, and quite close to the edge 
of the interferometer primary beam. In both images we also plot the 3$\sigma$ 
level of the two deuterated 
cores N and S shown in Fig.~\ref{fig_paperI}.

Because C3 and C4 are outside and on the edge of the primary beam, 
respectively, we decided not to discuss further these sources in the following.

Spitzer Post-Basic Calibrated Data (PBCD) at 24 and 70 $\mu$m were
obtained from the Spitzer Archive.  The data were taken in
Photometry/Super Resolution mode as part of program 20635 (R. Klein, PI).
Observations at 24 microns were obtained on 7 October 2005 (AORKEY
14944768), with an integration time of 99 s, and at 70 microns on 2 April
2006 (AORKEY 14945024) with an integration time of 76 s (using the Fine
Scale mode).  At present no IRAC observations have been made.
The left and right panels of Fig.~\ref{cont_spi_fig} show the 
Spitzer MIPS continuum images at 70 $\mu$m and 24 $\mu$m. In both 
images, two peaks roughly coincident with the
submillimeter continuum sources C1 and C3 are detected. At 70 $\mu$m,
an extended emission is detected towards C2, S and the southern portion
of N, but we believe that it is just due to the low angular resolution of
the map ($\sim 18$\asec ), which do not allow
to disentangle the contibution of the different sources.
The 24 $\mu$m continuum emission shows two peaks spatially coincident
to the 70$\mu$m ones, but they are both more centrally peaked than at
70 $\mu$m, and have smaller extension, probably due to the better angular resolution 
($\sim 6$\asec\ at 24 $\mu$m). A faint emission is detected towards C2, at the edge
of the main emission peak.
No emission is detected towards N and S at this wavelength.

\subsection{Molecular line emission.}

In paper I we have shown and discussed the distribution of the integrated 
intensity of the \D\  and  \H\  lines  (see also Fig.~\ref{fig_paperI}). However, as
said in Sect.~\ref{sma}, several transitions of other species have 
been observed and detected simultaneously to the  \D\  and  \H\ ones. 
In this Section, we complete the analysis started in paper I
and show a full report of the observations obtained.

\subsubsection{Integrated emission map of the optically thin component of the \H\ (1--0) line}
\label{n2hp}
 
\begin{figure}
\centerline{\includegraphics[angle=-90,width=8cm]{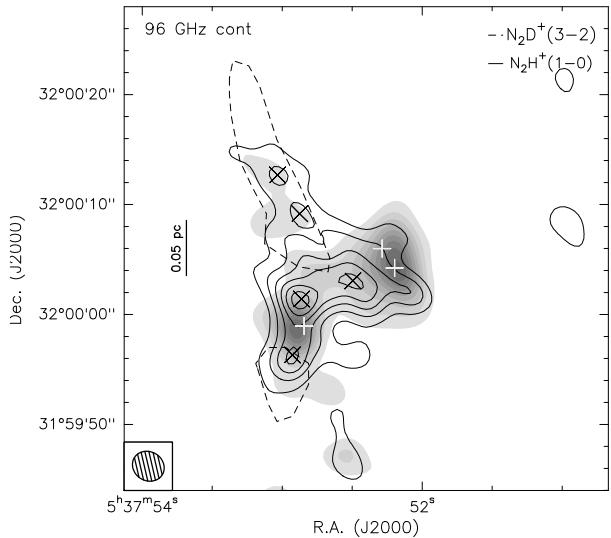}}
\caption{Integrated intensity of the hyperfine component $F_1 \,F = 1 0 \rightarrow 1 1 $
of the \H\ (1--0) line (solid contours), superimposed on the 96 GHz continuum
(grey-scale). The contours start from the 3$\sigma$ (0.018 Jy beam$^{-1}$), and are 
in steps of 3$\sigma$. The grey-scale is the same as in Fig.~\ref{cont_fig}. The
white crosses correspond to the three peaks detected in the 284 GHz continuum
image, and the black crosses mark the position of the \H\ emission peaks.
The dashed contrours represent the 3$\sigma$ level of the \D\ (3--2)
integrated emission. The ellipse in the bottom-left corner is the synthesised
beam.}
\label{map_thin}
\end{figure}

The integrated intensity of the main group of hyperfine components 
of the \H\ (1--0) line has been shown in Fig.~\ref{fig_paperI}. 
As already discussed in Sect.~\ref{Introduction}, the emission is extended and presents 
several bumps which are not resolved into separated emission peaks.
This may be the real shape of the emission in this tracer, or
it can be due to other reasons, such as an insufficient angular resolution, or
opacity effects. To better understand this, in Fig.~\ref{map_thin} we 
show the distribution of the integrated intensity of the  
hyperfine component $F_1 \,F = 1 0 \rightarrow 1 1 $ only, which is 
well separated from the other components and it is expected to be optically 
thin: from the fitting procedure described in Sect.~\ref{datared},
we obtain an optical depth for this component of $\sim 0.4\div 0.7$. Fig.~\ref{map_thin} 
shows that the extension of the integrated
intensity of this component is similar to that  shown in Fig.~\ref{fig_paperI}, 
but in this case we do detect five emission peaks. This suggests that, since the
angular resolution is the same, the lack of well separated
intensity peaks in Fig.~\ref{fig_paperI} is probably due to optical depth effects. 

The main emission peak is
located $\sim 4$\asec\ north of C2, while the others are
located about $\sim 5$\asec\ east of C1, and $\sim 5$\asec\ south
of C2, respectively, and two fainter peaks (detected at $\sim 6 \sigma$)
are seen towards core N. None of the millimeter sources 
coincide with the five \H\ emission peaks. These results indicate
that the source can host other dense cores which are not detected
in the millimeter continuum observations, probably because they are
too cold.

\subsubsection{Integrated emission map of the \CO\ (2--1) line wings }
\label{co}

Single-dish observations of the source in the \CO\ (2--1)
line have been performed by Zhang et al.~(\citeyear{zhang}), with
a $\sim 29$ \asec\ angular resolution. They detected a massive outflow
in the non-Gaussian line wings, the orientation of which is approximately in 
the WE direction, but the angular
resolution was not sufficient to detect neither the outflow center
nor the detailed morphology of the outflow itself.
The observations reported in this work significantly improve
the angular resolution of the previous ones, allowing us
to better determine the outflow shape and identify its origin. 

\begin{figure}
\centerline{\includegraphics[angle=0,width=8.0cm]{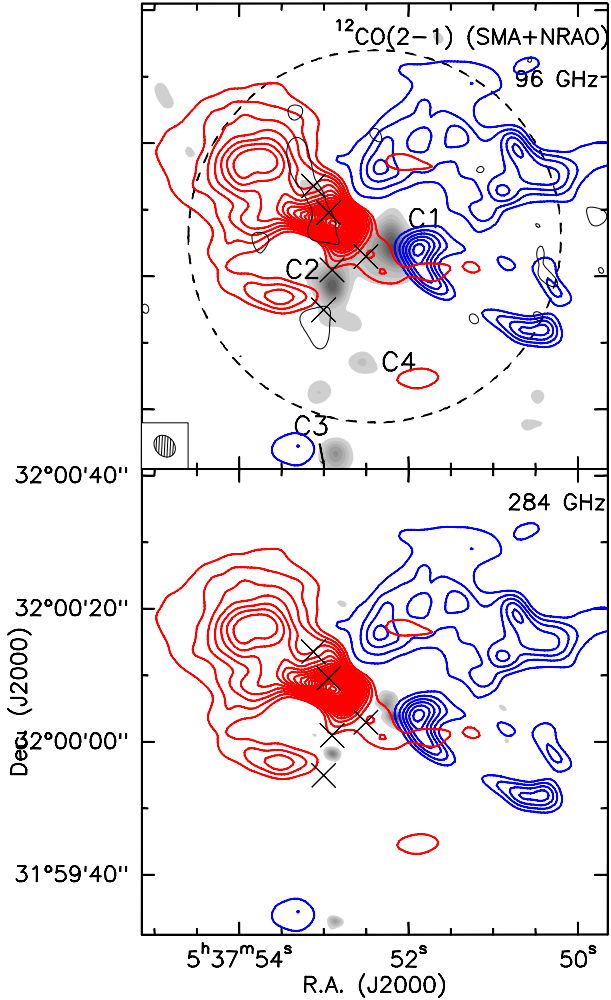}}
\caption{Top panel: red- and blue-shifted integrated emission
of the \CO\ (2--1) line (combined SMA + NRAO data), superimposed on the 96 GHz 
continuum map observed with the PdBI (grey-scale). Both red and blue contours 
start from 0.25 Jy beam$^{-1}$ ($\sim 3 \sigma$ rms), and are in steps of 0.25 
Jy beam$^{-1}$. The solid black contours represent the 3$\sigma$ level
of the \D\ (3--2) line integrated emission, and indicate the location 
of the deuterated cores N and S (see also Fig.~\ref{fig_paperI}). The crosses correspond to the
five \H\ (1--0) emission peaks shown in Fig.~\ref{map_thin}.
The dashed circle represents the SMA primary beam at the frequency
of the \CO\ (2--1) line. 
The beam of the combined \CO\ map is shown in the bottom left corner.
Bottom panel: same as top panel but the grey-scale corresponds to
the continuum at 284 GHz observed with SMA. The 3$\sigma$ level 
of the \D\ (3--2) emission and the interferometer primary beam are not shown here.
}
\label{cooutflow}
\end{figure}

In Fig.~\ref{cooutflow}, we show the map of the integrated intensity 
in the \CO\ (2--1) line wings, derived from channel maps obtained 
combining SMA and NRAO data, as described in Sect.~\ref{datacomb}. 
The blue- and red-shifted emissions have 
been averaged in the velocity intervals (--46.4, --26) \kms\ and (--9.2, 6.4) \kms , 
respectively. The outflow axis is predominantly oriented in the 
WE direction, with redshifted gas in the east and blueshifted
gas in the west. The lobes are clearly separated and have
approximately a biconical shape. The outflow center is near the 
position of the continuum source C1, and the exciting source
can be either C1-a and C1-b. The source coincident with the most western
peak of the \H\ (1--0) line integrated emission can also contribute
to the observed emission, but less likely than the continuum sources,
due to the fact that there are no other signs of protostellar activity
associated with it. The blueshifted
gas shows a fainter secondary peak to the north
of the map. 
From geometrical considerations, this 
northern blue-shifted emission might be driven by a source
within N, rather than C1-a or C1-b. However, as we will further
discuss later, the sources eventually
embedded within N are probably in the pre--stellar phase, i.e. prior to
 the main accretion phase in which the outflow is expected to form,
 so that this solution seems to us very unlikely.

The outflow length, from end-to-end, is approximately 35\asec\
(if we do not consider the secondary peaks of the blueshifted
emission), corresponding to $\sim 0.28$ pc at a distance of 1.8 kpc.
The semi-opening angle is between about 30 and 40$\degr$, and the
spatial separation of the lobes suggests that the inclination angle
with respect to the line of sight is likely to be very close
to the plane of the sky (see also Cabrit \& Bertout~\citeyear{ceb}).

In Table~\ref{tab_wings} we give the line parameters for the
calculation of the outflow properties, namely: 
 the velocity range of the blue- and red- wings, 
$\Delta V_{\rm b}$ and $\Delta V_{\rm r}$ (Cols.~1 and 2, respectively), the 
integrated intensity of \CO\ (2--1) in the wings, $\int T{\rm d}V$ (blue) and (red)
(Cols.~3 and 4; $T$ is the main beam brigthness temperature), and the maximum 
velocity of the wings, $V_{\rm max_{b}}$ and 
$V_{\rm max_{r}}$ (Cols.~5 and 6), defined as the difference
between the maximum velocity of the blue- and red- wings and the 
systemic velocity (i.e. $-18.4$ \kms ). The values listed in Cols.~3 and 4
are averaged values over the blue and the red lobes, 
respectively.

\begin{table*}
\begin{center}
\caption[] {Line parameters for calculations of the \CO\ (2--1) outflow. }
\label{tab_wings}
\begin{tabular}{cccccc}
\hline \hline
\small
 $\Delta V_{\rm b}$ & $\Delta V_{\rm r}$ & $\int T{\rm d}V$ (blue) & $\int T{\rm d}V$ (red) & $V_{\rm max_{b}}$ & $V_{\rm max_{r}}$ \\
 (\kms\ )  &  (\kms\ ) & (K \kms\ ) & (K \kms\ )  & (\kms\ )  &(\kms\ )   \\
\hline
--46.4, --26 & --9.2, 6.4 & 1329.3 & 1663.4 & 28 & 24.8 \\
\hline
\end{tabular}
\end{center}
\end{table*}

\subsubsection{Integrated intensity of other lines} 
\label{others}

\begin{figure*}
\centerline{\includegraphics[angle=-90,width=15cm]{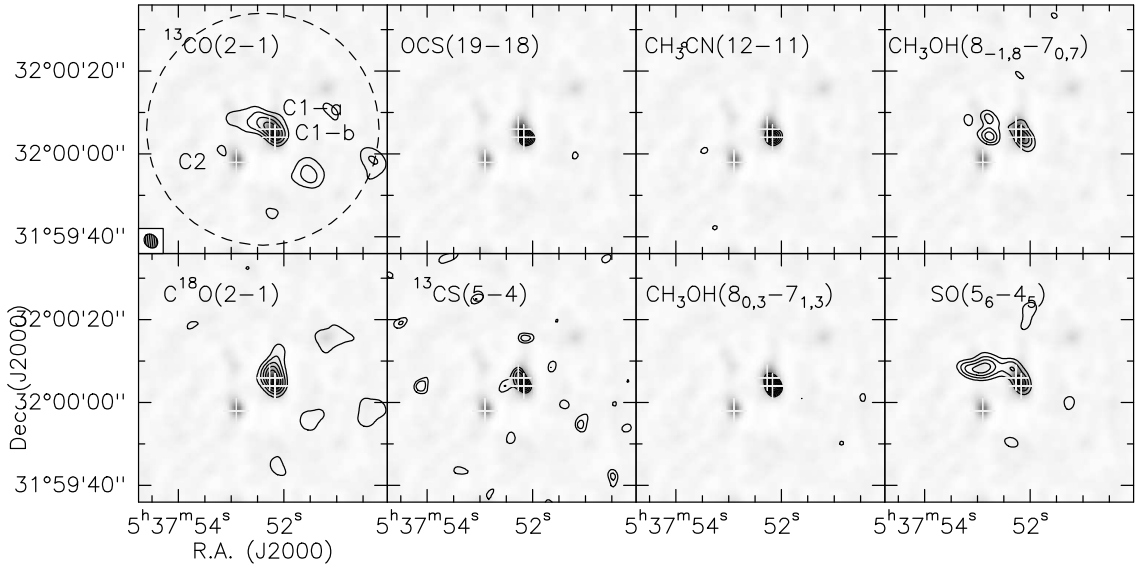}}
\caption{Integrated maps of all the molecular lines listed
in Table~\ref{tab_mol}, detected towards
\i\ with the SMA. For all transitions, the emission has been averaged over all the
channels with signal. In each panel, the transition is indicated in the
top left corner, the grey scale represents the 94 GHz continuum, and
the position of the mm cores is indicated by the white crosses.
For \CI , \CII , CH$_3$OH (8$_{-1,8}$--7$_{0,7}$) and SO, the contours start 
from the 3$\sigma$ level of the averaged map, and are in steps of 3$\sigma$, 
while for the other lines are in step of 1$\sigma$.
In the top left panel, where the \CI\ (2--1) line is shown,
the ellipse in the bottom left corner indicates the
synthesised beam, and the dashed circle corresponds to the interferometer 
primary beam. The other lines have the same primary beam and 
similar synthesised beams.
}
\label {figura_coetal}
\end{figure*}

In Fig.~\ref{figura_coetal} we show the distribution of the integrated
intensity of all the detected transitions listed in Table~\ref{tab_mol}, 
averaged over the channels with signal, but \CO\ (2--1) that has
been already discussed. 

The integrated intensity of the \CI\ (2--1) line peaks at the position of C1, but it shows an
elongated shape towards the north-east, suggesting that some of the 
emission comes from the outflow. On the other hand, the \CII\ (2--1) line emission is
less extended and it seems to arise from a common gaseous envelope in which
C1-a and C1-b are embedded. No significant emission of this line is detected
towards the \CO\ outflow lobes. A similar distribution is seen in the
integrated emission of $^{13}$CS (5--4) and, a bit less, in 
CH$_3$OH (8$_{0,3}$--7$_{1,3}$), showing that these transitions
trace the same gas.

The integrated intensity of the two high excitation lines
OCS (19--18) and CH$_3$CN (12--11) clearly 
shows the presence of a compact hot-core centered on source C1-b. 
In both tracers, the hot core is unresolved, so that its diameter has an 
upper limit of approximately 3\asec , i.e. $\sim 0.026$ pc at the given 
distance. The hot-core nature of this condensation is confirmed
by the gas temperature of $\sim 200$ K that we will derive in Sect.
~\ref{ch3cn}.

For the CH$_3$OH (8$_{-1,8}$--7$_{0,7}$) and SO (5$_6 - 4_5$) line, 
the emission looks clumpy, with a main spot corresponding to the
position of C1, and another prominent spot coincident with the 
red-shifted \CO\ emission, and several fainter spots north and west 
of C1 which do not coincide with none of the other tracers observed.
The strong emission of SO (5$_6 - 4_5$) in the red lobe of the \CO\ outflow is consistent
with previous studies, in which it has been found that SO abundance is
enhanced in outflows (Bachiller et al.~\citeyear{bachiller}, Viti et al.~\citeyear{viti}),
and it has been clearly detected in other massive outflows studied at
high-angular reoslution (see e.g.~the case study of AFGL 5142, Zhang et
al.~\citeyear{zhang07}).

\begin{figure}
\centerline{\includegraphics[angle=0,width=8cm]{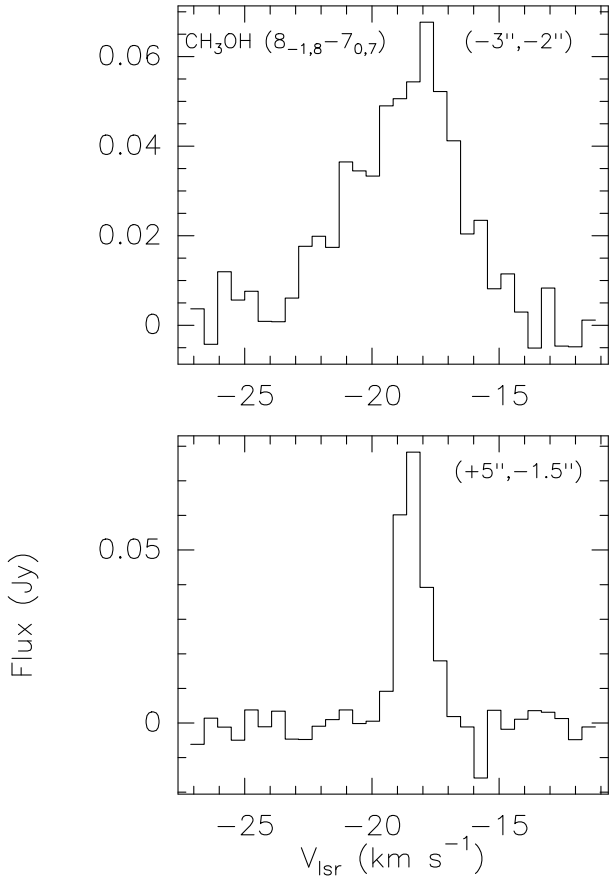}}
\caption{Spectra of the CH$_3$OH(8$_{-1,8}$--7$_{0,7}$) line
obtained towards the positions of the two main emission peaks seen in the
averaged map of this line (Fig.~\ref{figura_coetal}). The offsets (in
arcsec) are indicated in the top-right corner, and correspond to the peak
position of C1 and the 'base' of the red lobe of the \CO\ outflow.
}
\label {ch3oh_spec}
\end{figure}

On the other hand, the CH$_3$OH (8$_{-1,8}$--7$_{0,7}$) line is believed
to be a Class I methanol maser line, and it has been detected
at low-angular resolution in several high-mass star forming regions (see 
e.g. Slysh et al.~\citeyear{slysh}). Recently, the SMA has imaged
several masers in this line towards the source DR21(OH) (Bourke, priv.
comm.). 
Class I methanol masers are often located offset from the position of hot cores
or HII regions, where Class II and other masers, such as those of water and OH, 
are found. They are believed to be collisionally excited, probably where a 
powerful outflow impacts the surrounding quiescent material (see e.g.~Sandell 
et al.~\citeyear{sandell}). However, high-angular 
resolution observations have shown that in some cases they
can be found in close proximity of HII regions and water masers (Kurtz
et al.~\citeyear{kurtz04}).
The map shown in Fig.~\ref{figura_coetal} clearly show
that the (8$_{-1,8}$--7$_{0,7}$) line comes from both the hot core and 
the red lobe of the outflow, indicating that it is originated
by shocked gas, and the spatial extent is similar to that of SO, even 
though the SO emission is more prominent in the outflow
than that of CH$_3$OH.  To better understand the nature
of the CH$_3$OH (8$_{-1,8}$--7$_{0,7}$) line, in Fig.~\ref{ch3oh_spec} we
show the spectra of this line
towards two positions at an offset of (--3\asec , --2\asec ) and
(+5\asec , --1.5\asec ) from the map centre. These two positions correspond
to the main emission peaks of the averaged map, and roughly coincide 
with C1 and with the inner part of the red lobe of the \CO\ 
outflow, respectively (see Fig.~\ref{figura_coetal}). We clearly see that
the line in the outflow is much narrower than the one observed towards
core C1 ($\sim 5$ \kms\ and $\sim 1$ \kms , respectively). The broad line 
seen towards C1 is likely thermal, while the narrow 
one observed towards the red lobe of the \CO\ outflow could come from a 
single maser spot, even though the flux density is not very high ($\sim 0.08$
Jy) and could also be thermal as well.

\begin{figure}
\centerline{\includegraphics[angle=0,width=8cm]{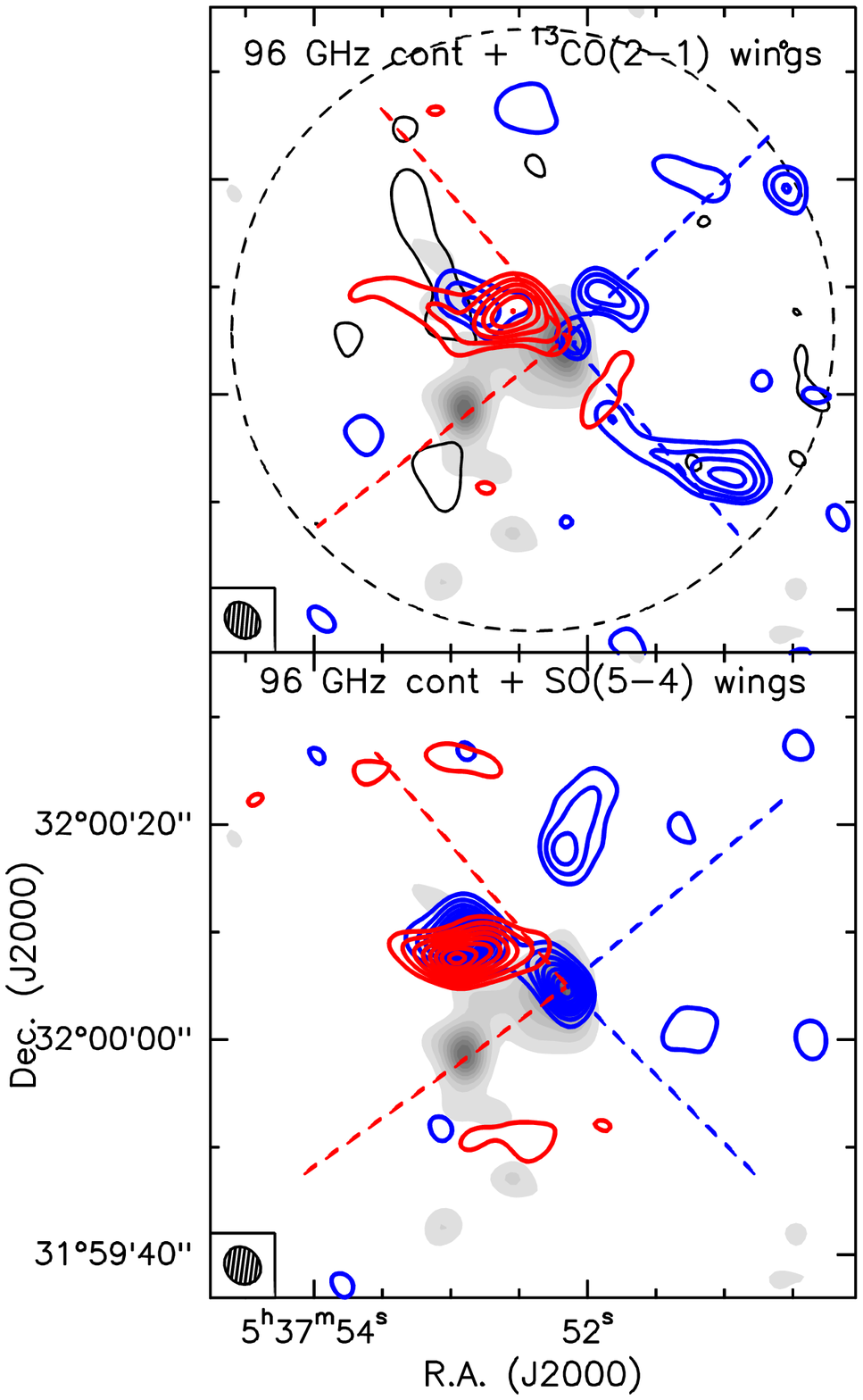}}
\caption{Top panel: red- and blue-shifted integrated emission
of the \CI\ (2--1) line observed with SMA, superimposed on the 96 GHz 
continuum map observed with the PdBI (grey-scale). For this latter
we use the same contours
as in Fig.~\ref{cooutflow}. The integration ranges in velocity are (--15.88; --8.14)
and (--23.06; -20.85) \kms . The red contours start from 0.16 
Jy beam$^{-1}$ ($\sim 3 \sigma$ rms), and are in steps of 0.16 
Jy beam$^{-1}$. The blue contours start from 0.24 Jy beam$^{-1}$ 
($\sim 3 \sigma$ rms), and are in steps of 0.16 Jy beam$^{-1}$
The solid black contours represent the 3$\sigma$ level
of the \D\ (3--2) line integrated emission as in Fig.~\ref{cooutflow}. 
The dashed circle corresponds to the SMA primary beam at the frequency
of the \CI\ (2--1) line. The red- and blue-dashed lines
indicate the opening angle of the \CO\ outflow lobes as 
in Fig.~\ref{cooutflow} (excluding the northern blue bump). 
The SMA synthesised beam is shown in the bottom left corner.
Bottom panel: same as top panel for the SO (5$_6 - 4_5$) line. 
The integration ranges in velocity are (--23.5; --20.2) and (--14;--3)
\kms . For both red- and blue-shifted emission, both the first contour and the step
is the 3$\sigma$ of the averaged map (0.06 and 0.09 Jy beam$^{-1}$,
respectively). The 3$\sigma$ level 
of the \D\ (3--2) emission and the interferometer primary beam are not shown here.
}
\label{fig_13co}
\end{figure}

Finally, to better highlight the emission of \CI\ and SO in the outflow,
in Fig.~\ref{fig_13co} we show the integrated intensity maps of the
blue- and red-shifted emission of the \CI\ (2--1) and SO (5$_6 - 4_5$) lines. 
The blue-shifted emission of \CI\ seems to trace the cavity created 
by the blue lobe of the \CO\ outflow, while the red-shifted emission follows 
the inner part of the \CO\ red lobe (see upper panel of Fig.~\ref{fig_13co}). 
For SO (5$_6 - 4_5$), we find a similar result for the red-shifted
emission, while the blue-shifted emission of this line does not follow that
of  \CO , but it arises mostly from core C1 and from the \CO\ red lobe. This
strange feature could be explained by the presence of another outflow not detected in \CO ,
centered roughly in between core N and core C1.
A northern bump in the blue-shifted emission is also detected, located at the position of
 the northern blue-shifted emission seen in Fig.~\ref{cooutflow}. 

\section{Derivation of the physical parameters}
\label{phy_par}

\subsection{Temperature of core C1-b from CH$_3$CN}
\label{ch3cn}

\begin{figure}
\centerline{\includegraphics[angle=0,width=8cm]{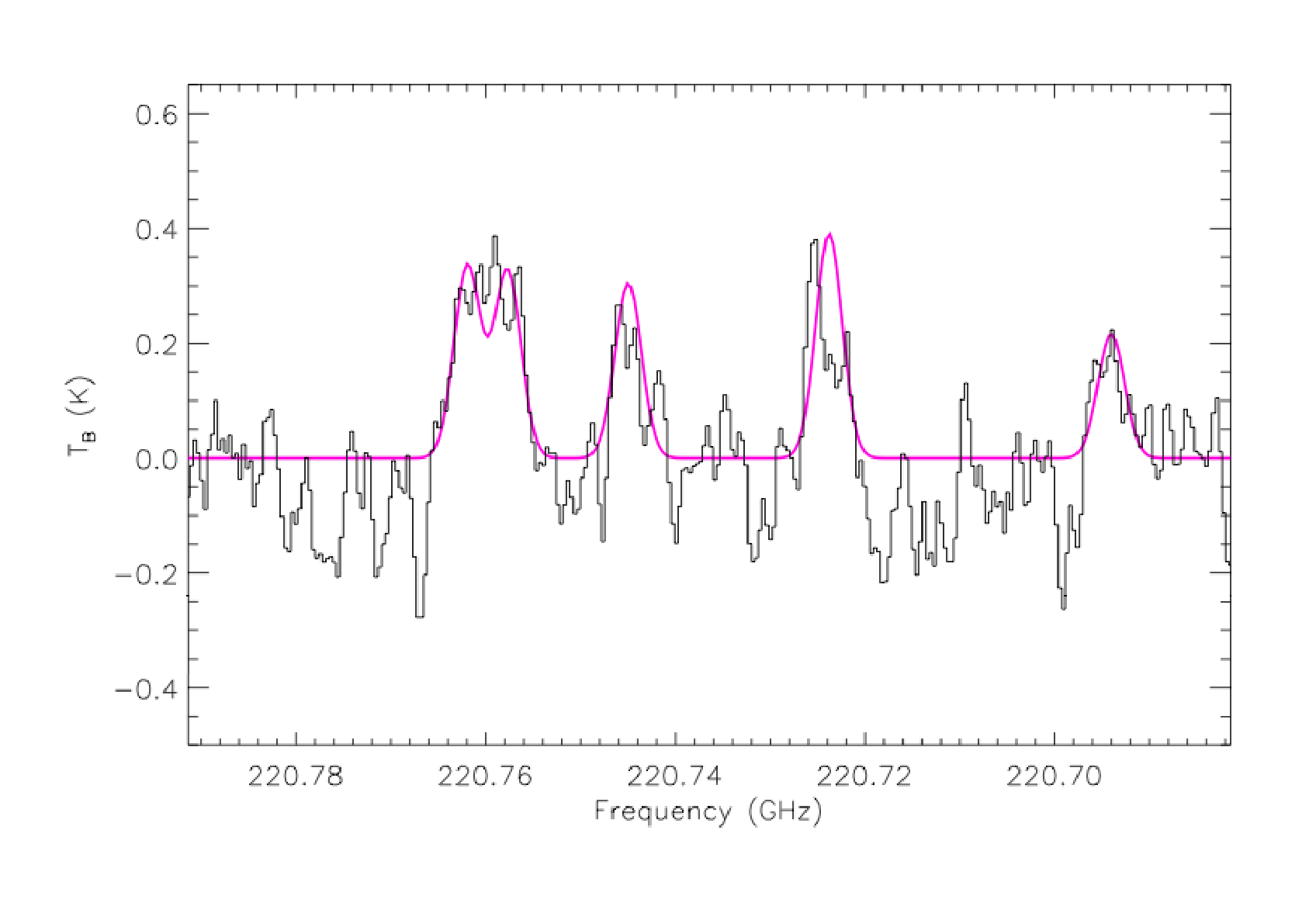}}
\caption{Spectrum of the CH$_3$CN (12--11) line at the peak position, obtained with the SMA. The
red line represents the best fit to the components $K=0$ to $K=4$, performed as explained
in Sect.~\ref{ch3cn}.
}
\label{fig_ch3cn}
\end{figure}

Figure~\ref{fig_ch3cn} shows the CH$_3$CN (12--11) spectrum at the peak position
of the emission map shown in Fig.~\ref{figura_coetal}. CH$_3$CN is a symmetric-top molecule 
whose rotational levels are described by two quantum numbers: J, associated with the total angular 
momentum, and K, its projection on the symmetry axis. Such a structure entails that for each radiative
transition, $J+1\rightarrow J$, lines with $K\leq J$ can be seen (for a detailed description see 
Townes \& Schawlow 1975). In our observations the bandwidth covers up to the $K=7$ component 
for the (12--11) transition. However, we detected only lines up to K=4, for which the energy of
the upper level is 184 K.
In order to compute the line parameters, we have performed Gaussian fits to the observed spectrum assuming that all 
the K components arise from the same gas. Hence, they have the same LSR velocity and line width. 
We fixed the line separation in the spectrum to the laboratory value 
and derive a line width by fitting the spectrum of all the K components. Then, under 
the assumption of LTE conditions, we compute the CH$_3$CN spectrum using the gas temperature, 
the CH$_3$CN density, the source size as input parameters, and a line width fixed to the observed 
value (Qiu 2009, private communication). This approach does not assume optically thin emission 
in CH$_3$CN, as does in the Boltzmann analysis (see e.g. Kuiper et al. 1984; Bergin et al. 1994). 
%
The best fit to the CH$_3$CN spectrum shown in Fig.~\ref{fig_ch3cn} yields a  range of 
kinetic temperatures from 150 - 250 K. These values are comparable to typical temperatures found
in hot molecular cores surrounding a newly formed massive star
(see e.g.~Kurtz et al.~\citeyear{kurtz}).

\subsection{Physical parameters from dust emission}
\label{par_dust}

\begin{table*}
\tiny
\caption{Peak position, angular and linear diameter, integrated flux density, mass, H$_2$ volume 
and column density of the millimeter condensations C1 (resolved into C1-a and C1-b in the 284 GHz image) and 
C2. The masses are computed for $\beta = 2$, and assuming $T$ = 30K. The H$_2$ volume and
column densities are calculated assuming a spherical source with diameter equal to the deconvolved
3$\sigma$ level.}
\label{cont_tab}
\begin{center}
\begin{tabular}{ccccccccc}
\hline \hline
\multicolumn{9}{c}{$\nu = 96$ GHz} \\
Source & \multicolumn{2}{c}{Peak position} & \multicolumn{2}{c}{Diameter}  & $F_{\nu}$ & $M_{\rm cont}$ & $n_{\rm H_2}$  & $N({\rm H_2})$ \\
\cline{2-3} \cline{4-9} 
       & R.A.(J2000) & Dec.(J2000)  & $\theta_{\rm s}$  & $D$  & &  &  & \\
       & 05$^h$37$^m$ &   & (\asec)& (pc) & (mJy) & (M$_{\odot}$) & ($\times 10^6$ \cmc ) & ($\times 10^{23}$ \cmq )  \\
C1$^{\bf *}$ & 52$^s$.17 & +32$ \degr$00$^{\prime}$ 04\pas 5 & 9.08 & 0.08  &  11 & 27 & 2.1 &  5.2 \\
C2 & 52$^s$.91 & +31$ \degr$59$^{\prime}$ 58\pas 5 & 6.35 & 0.06  &  6  & 15 & 2.8 &  5.2 \\
\hline
\multicolumn{9}{c}{$\nu = 225$ GHz} \\
Source & \multicolumn{2}{c}{Peak position}  & \multicolumn{2}{c}{Diameter}  & $F_{\nu}$ & $M_{\rm cont}$ & $n_{\rm H_2}$ &  $N({\rm H_2})$ \\
\cline{2-3} \cline{4-9} 
       & R.A.(J2000) & Dec.(J2000)  & $\theta_{\rm s}$  & $D$  & & & & \\
       & 05$^h$37$^m$ &   & (\asec)& (pc) & (mJy) & (M$_{\odot}$) & ($\times 10^6$ \cmc ) & ($\times 10^{23}$ \cmq )  \\
C1$^{\bf *}$ & 52$^s$.19 & +32$ \degr$00$^{\prime}$ 05 & 7.3 & 0.06  &  150 & 12 & 1.4 &  3.0 \\
C2 & 52$^s$.90 & +31$ \degr$59$^{\prime}$ 58\pas 5 & 5.3 & 0.05  &  60  & 4.7 & 1.2 &  2.0 \\
\hline
\multicolumn{9}{c}{$\nu = 284$ GHz} \\
Source & \multicolumn{2}{c}{Peak position}  & \multicolumn{2}{c}{Diameter}  & $F_{\nu}$ & $M_{\rm cont}$ & $n_{\rm H_2}$ &  $N({\rm H_2})$ \\
\cline{2-3} \cline{4-9} 
       & R.A.(J2000) & Dec.(J2000)  & $\theta_{\rm s}$  & $D$  & & & & \\
       & 05$^h$37$^m$ &   & (\asec)& (pc) & (mJy) & (M$_{\odot}$) & ($\times 10^6$ \cmc ) & ($\times 10^{23}$ \cmq )  \\
C1-b & 52$^s$.17 & +32$ \degr$00$^{\prime}$ 04 & 1.8 & 0.016  &  160 & 5.2;  0.6$^{\dagger}$ & 8.5;  1.0 $^{\dagger}$ & 7.6; 0.9$^{\dagger}$  \\
C1-a & 52$^s$.27 & +32$ \degr$00$^{\prime}$ 06 & 2.1 & 0.018  &  115  & 3.6 & 6.6 &  5.7 \\
C2  & 52$^s$.90 & +31$ \degr$59$^{\prime}$ 58 & 1.5 & 0.013  &  90  & 2.9 & 10.9 &  7.4 \\
\hline
\end{tabular}
\end{center}
\footnotesize
$^{\bf *}$ = C1-a + C1-b \\
$^{\dagger}$ = computed from the CH$_3$CN temperature of 200 K \\
\normalsize
\end{table*}

The physical parameters derived from the millimeter continuum images
presented in Sect.~\ref{cont} are listed in Table~\ref{cont_tab}:
the position of the emission peak, in R.A. (J2000) and Dec. (J2000),
of C1 (resolved into C1-a and C1-b at 284 GHz) and C2
are listed in Cols.~2 and 3, respectively; the deconvolved angular ($\theta$) 
and linear ($D$) diameters are given in Cols.~4 and 5. 
Assuming that the emission has a Gaussian profile, we have 
derived the angular diameter of the sources deconvolving
the observed FWHM with a Gaussian corresponding to the 
synthesised beam of the interferometer. 
The observed FWHM has been measured as the geometrical mean of major 
and minor axes of the contour at half of the intensity peak.
The linear diameters have been computed using a source distance of 
1.8~kpc (Zhang et al.~\citeyear{zhang}).
  
In Cols.~6 of Table~\ref{cont_tab} we also give the flux density, 
$F_{\nu}$, derived integrating over the 
3$\sigma$ level in each core. Col.~7 shows the gas masses obtained 
from $F_{\nu}$: assuming constant gas-to-dust ratio, optically thin and 
isothermal conditions, the total gas+dust mass is given by:
\begin{equation}
M_{\rm cont}=\frac{F_{\nu} d^2}{k_{\nu}B_{\nu}(T)}\;.
\label{eqdustmass}
\end{equation}
In Eq.~(\ref{eqdustmass}), $d$ is the distance, $k_{\nu}$ is the dust 
opacity coefficient, derived according to
$k_{\nu}=k_{\nu_0}(\nu/\nu_0)^{\beta}$ (where $\nu_0=230$ GHz and 
$k_{\nu_0}=0.005$ cm$^2$ g$^{-1}$, which implies a gas-to-dust ratio of 
100, Kramer et al.~\citeyear{kramer}), and
$B_{\nu}(T)$ is the Planck function calculated at the dust temperature 
$T$. For this latter, since we do not have a direct estimate (except for C1-b),
we have assumed a 'reasonable' value of 30 K (see e.g.~Molinari
et al.~\citeyear{mol00}).  For C1-b, for which we also have a direct
temperature measurement from CH$_3$CN (see Sect.~\ref{ch3cn}), we give 
two estimates: one assuming $T =30$ K and the other assuming the
value derived from the CH$_3$CN spectrum, i.e. $T=200$ K.
For the dust grain emissivity index, we 
assumed $\beta = 2$ (see e.g. Hill et al.~\citeyear{hill}). 
From the 3$\sigma$ level in the continuum images, we have also derived that
we are sensitive to point-like sources with masses of 0.6, 0.2 and 0.01\solm\ at
284, 225 and 96 GHz, respectively (assuming $T=30$ K).
We also estimate a column density sensitivity of $\sim 7 \times 10^{22}$, 
$1.9 \times 10^{22}$ and $1.4 \times 10^{21}$ \cmq\ at the
three frequencies mentioned above.

In Cols.~8 and 9 of Table~\ref{cont_tab}, we also list the average H$_2$ volume and
column densities. The H$_2$ average volume densities have been computed assuming 
spherical cores, then the column densities have been derived by multiplying the 
volume densities for the core diameters. We have used 
as angular diameters those obtained deconvolving the 3$\sigma$ levels in the continuum 
images, and not those listed in Table~\ref{cont_tab} because the continuum fluxes have been
derived integrating the emission within the 3$\sigma$ contours. All volume densities are of the
order of $10^{6}$ \cmc , while the column
densities are of the order of $10^{23}$ \cmq , which are typical values found 
in cores associated with intermediate-/high-mass forming stars, and are higher
than the critical value of 1 gr \cmq\ required to form a massive star at core centre
(Krumholz et al.~\citeyear{kru}), assuming that the gas is totally made of H$_2$.

The mass of C2, estimated to be 11.8, 4.7 and $\sim 2.9$ \solm\ at
96, 225 and 284 GHz, indicates that the core probably hosts an
intermediate-mass object. The mass of C1 is 27 and 11.8 \solm\ from the 96 and 
225 GHz continuum emission, and when it is resolved in two components, C1-a and 
C1-b, in the 284 GHz image, the masses of the components are 3.8 and 
5.2 \solm , respectively. These values are typical of cores associated with intermediate-mass 
protostellar objects (see e.g. Beltr\`an et al.~\citeyear{beltran08}). 
For C1-b, adopting the temperature of 200 K from CH$_3$CN, we even obtain 0.6 \solm . 
However, as we will discuss better in Sect.~\ref{nat_cont}, the source embedded in core C1-b
is likely an early-B newly formed star. Therefore, the gas mass
derived from the millimeter continuum represents for this source the mass of the 
circumstellar material only, and not that of the central star. 

Additionally, for all sources these mass estimates, especially
those derived from the 284 GHz continuum, are
very uncertain because affected by several problems. First, both the
dust temperature and $\beta$ assumed are very uncertain: $\beta$ 
is expected to be between 1 and 2 in high-mass star forming regions
(see e.g.~Hill et al.~\citeyear{hill} and references therein),
while the dust temperature can vary considerably from a source 
to another (see e.g. Molinari et al.~\citeyear{mol00}; Sridharan
et al.~\citeyear{sridharan}). 
For example, assuming the gas temperature derived 
from \AMM\ observations, that is $T=17$K (Jijina et al.~\citeyear{jijina}), 
with $\beta =2$ we get 7.9 and 10.9 \solm\ for C1-a and C1-b, respectively, 
and 8.8 and 12.2 assuming $T=17$K and $\beta=1.5$ (Mathis \& 
Wiffen~\citeyear{mew}). 
Second, the 284 GHz image is more affected than the others by
the problem of the missing flux, so that the core masses derived
from this image are representative of the circumstellar material very close 
to the central object only, while those 
from the 225 and 96 GHz continuum are more representative of the whole
envelope. To estimate the amount of
flux filtered out beacuse of missing short spacing information, we have compared
the flux measured with the single-dish SCUBA map at 850 $\mu$m 
(Fontani et al.~\citeyear{fonta06}) to that measured in the 284 GHz map:
the peak flux of the single-dish map is $\sim 2.0$ Jy beam$^{-1}$. Assuming
a spectral index of 4, the peak intensity at 1 mm (i.e. 284 GHz)  is 1.04 Jy beam$^{-1}$.
After smoothing the interferometric map at the SCUBA angular resolution of
$14$ \asec , the flux density in the area corresponding to the
SCUBA beam is 0.11 Jy beam$^{-1}$, which means that we are recovering
only $\sim 11\%$ of the total flux.
Therefore, the masses derived from the 284 GHz continuum maps have to be 
considered as lower limits for the circumstellar masses. On the other hand,
at 96 GHz, the expected flux is 0.011 Jy beam$^{-1}$ while the value measured
from the PdBI map is 0.0075 Jy beam$^{-1}$, so that at this frequency we are
recovering 68$\%$ of the total flux.

\subsection{Physical parameters from \H\ and \D .}

\subsubsection{Deuterium fractionation.}
\label{par_n2hp}

For completeness, in Table~\ref{tab_dfrac}, we give the
\H\ and \D\ total column densities, $N$(\H ) and
$N$(\D ), and the Deuterium fractionation, \Dfrac ,
that we have derived in paper I. $N$(\H ), $N$(\D ) and 
\Dfrac = $N$(\D )/$N$(\H ),
derived for N and S, are listed in Cols. 2, 3 and 4,
respectively, of the same Table. For more details on the 
methods used as well as in the assumptions made, see 
Sect.~3.1 of paper I. 
As already discussed, in both condensations, \Dfrac\ is $0.11$, 
which is comparable to the values of \Dfrac\ found in low-mass 
pre--stellar cores 
by Crapsi et al.~(2005), following the same method. 
These values are also comparable to those derived by
Pillai et al.~(2007) in infrared-dark clouds from deuterated 
ammonia. However, their observations are related to the very
cold, pc-scale molecular envelope, and not to compact sub-pc 
scale cores. 

\begin{table}
\caption{\H\ and \D\ total column densities ($N$(\H )
and $N$(\D )), deuterium fractionation (\Dfrac ), linear
diameter ($L$), and mass derived from $N$(\H ) ($M_{\rm N_2H^+}$)
for the \D\ condensations N and S. The uncertainties computed following
the standard propagation of the errors are given between parentheses.}
\label{tab_dfrac}
\begin{center}
\begin{tabular}{cccccc}
\hline \hline
source & $N({\rm N_2H^+})$ & $N({\rm N_2D^+})$ & $D_{\rm frac}$ & $L$ & $M_{\rm N_2H^+}$ \\
       & ($\times 10^{13}$\cmq )  & ($\times 10^{12}$\cmq )  & & (pc) & ($M_{\odot}$) \\
\hline
N    &  1.9(0.6) &  2.1(0.3) &    0.11(0.04) & 0.09 & 8.7 \\
S  &   1.5(0.5) &  1.6(0.3)  &    0.11(0.04) & 0.05 & 2.5 \\
\hline
\end{tabular}
\end{center}
\end{table}

\begin{figure}
\centerline{\includegraphics[angle=0,width=7cm]{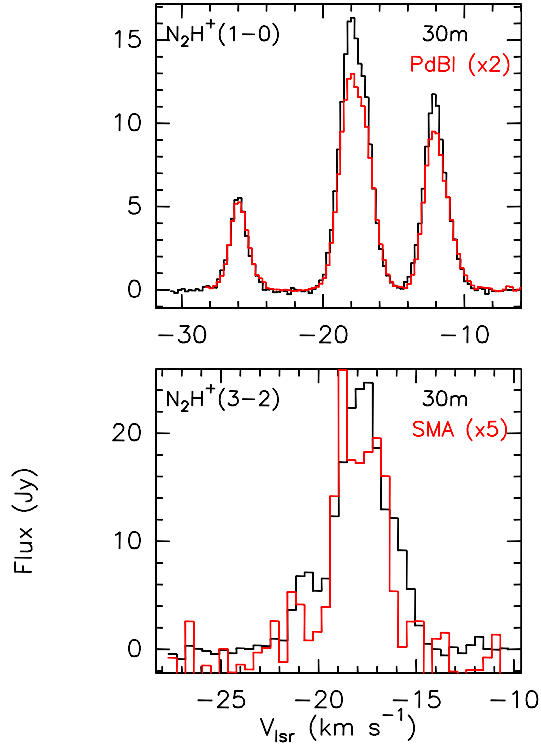}}
\caption{Flux density comparison between IRAM-30m spectra
and interferometric spectra obtained in the \H\ (1--0) (top panel)
and \H\ (3--2) (bottom panel) lines. The interferometric flux density 
of the \H\ (1--0) line, measured with the PdBI, has been multiplied by a
factor of 2, while that of the \H\ (3--2) line, measured with the SMA,
by a factor of 5. Both PdBI and SMA spectra have been obtained
integrating the maps in Figs.~1 and 2 of paper I
over the 3$\sigma$ level.} 
\label{spec_comp}
\end{figure}

We could not derive the column densities of N and S from the
\H\ (3--2) line because this is undetected towards the two
molecular condensations, as shown in Fig.~1 of paper I.
Such a different distribution of the integrated intensity 
with respect to that of the \H\ (1--0) line is
probably due to the extended emission being
filtered out differently by the two interferometers. To compute the
amount of the missing flux, we have compared our interferometric
spectra with the single-dish spectra obtained with the IRAM-30m telescope
(see Fontani et al.~2006). We have resampled the single-dish 
spectra of \H\ (1--0) and (3--2) to the same resolution in velocity 
as the interferometric spectra. 
The superposition of the spectra is shown in Fig.~\ref{spec_comp}: 
In the \H\ (1--0) line, the flux measured by the PdBI is $\sim 2$ times 
less than that measured with the IRAM-30m telescope, while in the
\H\ (3--2) line the flux measured with the SMA is only 
one fifth of that measured with the 30m antenna. This indicates
that the extended emission is much more resolved in
the SMA map of the (3--2) line than in the PdBI map of the (1--0) line.

\subsubsection{Velocity field.}
\label{par_kin}

From the molecular lines observed one can
derive information about the velocity field in the source.
In particular, the peak velocity, $V_{\rm LSR}$, of the lines gives information on ordered
motions of the gas (rotation, inward or outward motions), while the line 
widths, $\Delta V$ provide information on the gas turbulent motions.
In this section, we show the results obtained from the \H\ (1--0) line,
for which the spectra are not affected by central dips as those of 
\CO\ and its isotopologues. Also, this tracer is fairly well detected
towards both the continuum cores and the deuterated ones, so that
it allows us to investigate both ordered and turbulent motions across
the whole source.

To derive $V_{\rm LSR}$ and $\Delta V$, we have adopted the following procedure: 
from the channel maps of the \H\ (1--0) line, we have extracted a spectrum from each 
pixel inside the 3$\sigma$ level of the integrated emission (pixel 
size $\sim 1$\asec $\times \; 1$\asec ), 
and then fitted these spectra as described in Sect.~\ref{datared}. 

\begin{figure*}[t!]
 \begin{center}
 \resizebox{\textwidth}{!}{\includegraphics[angle=-90, width=10cm]{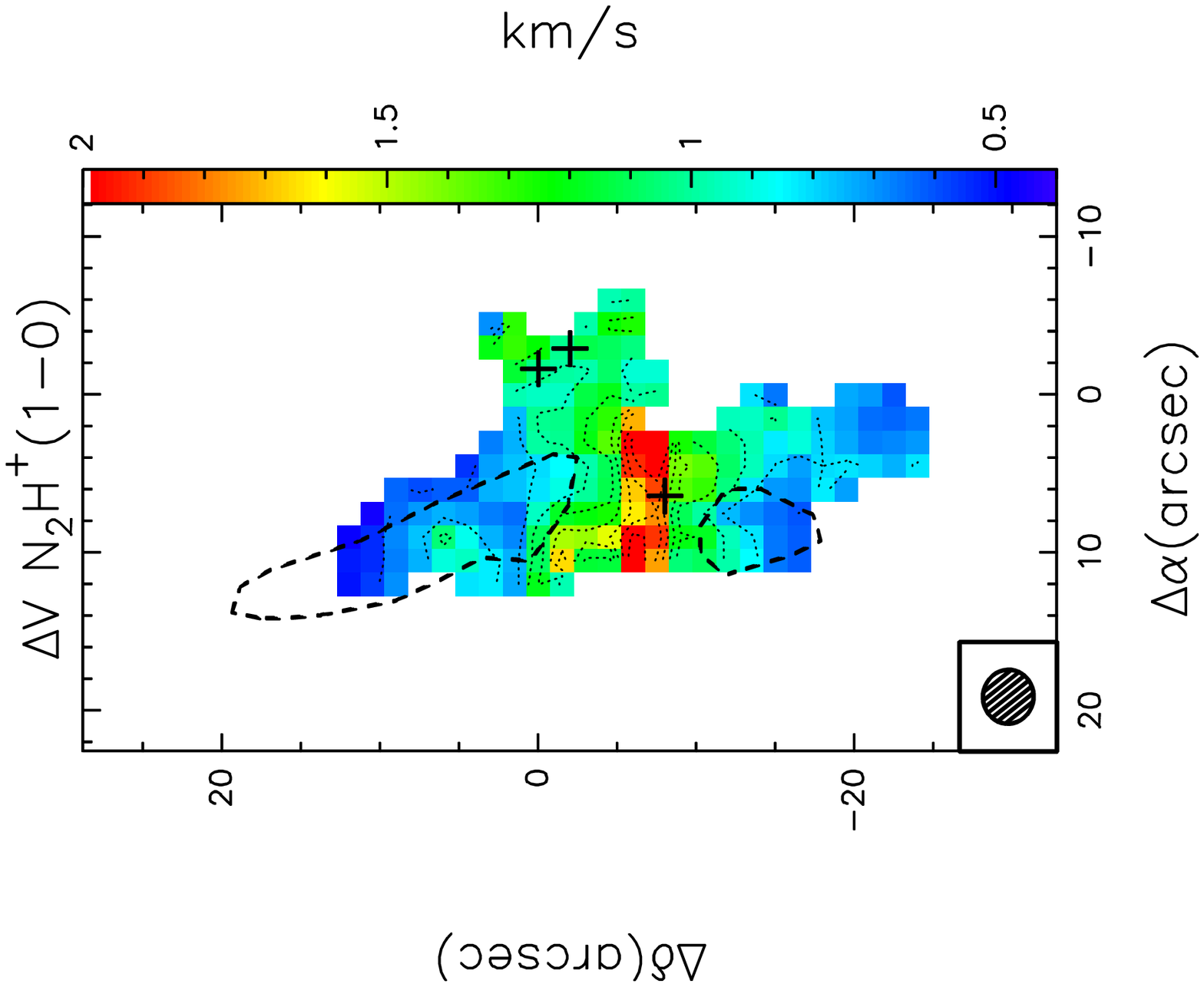}%
                           \includegraphics[angle=-90,width=10cm]{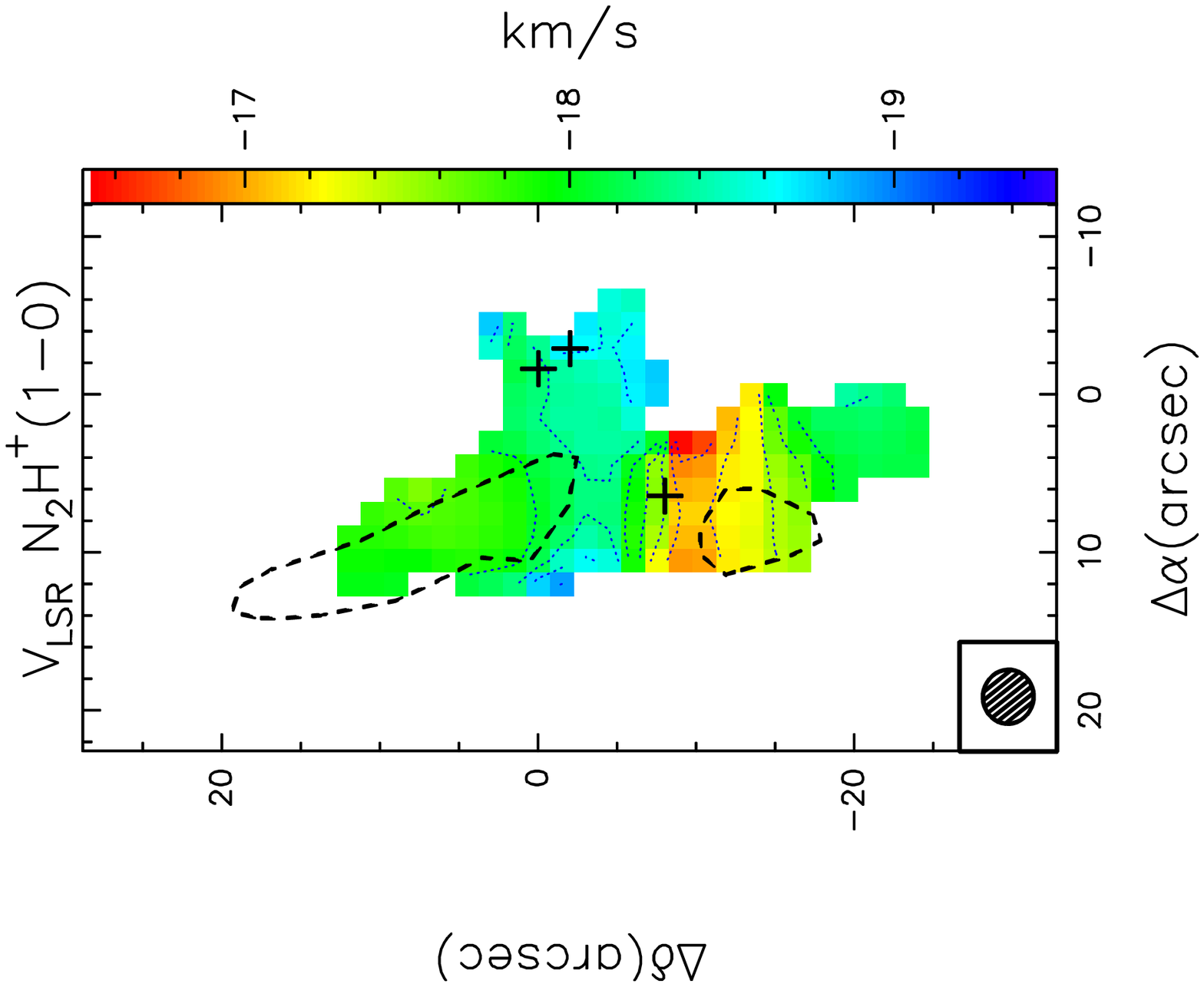}} 
 \end{center}
 \caption
{\label{fig_kintot} Left panel: Map of the \H\ (1--0) line width, $\Delta V$ 
(coloured-scale). The dotted contours range from 0.4 to 2.0 \kms ,
in steps of 0.2\kms .The crosses indicate the position of the continuum sources.  
The dashed lines represent the 3$\sigma$ level of N and S.
The PdBI synthesised beam at the frequency of the \H\ (1--0) line is
shown in the bottom left corner. 
Right panel: same as top panel for the peak velocity, $V_{\rm LSR}$, of the \H\ (1--0) line.
The dotted contours range from $-16.8$ to $-19.2$\kms , in steps of $0.3$ \kms . }
\end{figure*}

In Fig~\ref{fig_kintot} we show the maps of $\Delta V$ (left panel),
and $V_{\rm LSR}$ (right panel) of the \H\ (1--0) line.
From the left panel, one can
see that the line widths are clearly broader than $\sim 1$ \kms\
in the region where the continuum cores C1 and C2 and  
the \H\ (1--0) main peaks are located. 
On the other hand, the gas is more 
quiescent around this central region, and towards the position
of N and S ($\Delta V$ between $\sim 0.4$ and 1 \kms ). These 
results indicate that in the continuum cores, as well as in the \H\ 
main peaks, the star formation process is actively taking place,
while in the deuterated cores the gas is more quiescent
because they are still in the pre--stellar phase. Interestingly,
the plot in the right panel of Fig.~\ref{fig_kintot} indicates
that in the region where the turbulence is higher
the gas is also red-shifted with respect to the systemic
velocity, while no clear trend is seen in
the other portions of the source, in which the gas velocity is
quite close to the systemic one. 
This is probably caused by the interaction with the
red lobe of the \CO\ outflow, which is expanding in that
portion of the cluster.

In Sect.~\ref{nat_det}, we will discuss in some more details
this interaction, and how
 $\Delta V$ and $V_{\rm LSR}$ vary in the deuterated cores N 
 and S, by using also the \D\ (3--2) line. 

\subsection{Outflow parameters from \CO\ line wings}
\label{par_outflow}

\begin{table}
\begin{center}
\caption[] {Physical parameters of the outflow:
column densities of the blue- and the red- wing ($N_{\rm b}$ and 
$N_{\rm r}$), masses ($M_{\rm b}$, $M_{\rm r}$ and $M_{\rm out}$), 
momentum ($P_{\rm out}$), energy ($E$), dynamical timescale 
($t_{\rm dyn}$), mass entrainment rate ($\dot{M}_{\rm out}$), mechanical 
force ($\dot{P}_{\rm out}$) and mechanical luminosity ($L_{\rm m}$). 
All parameters have been calculated without 
correction for the inclination angle
(Col.~2), assuming the mean inclination angle of $57^{\circ}$
(Col.~3) and an extreme inclination angle of $80 \degr$ (Col.~4).}
\label{outflow_tab}
\begin{tabular}{cccc}
\hline \hline
\multicolumn{4}{c}{\CO\ (2--1)} \\
 parameter & no corr. & $\theta=57 \degr$ & $\theta=80 \degr$ \\
\hline
  $N_{\rm b}$ ($\times 10^{21}$cm$^{-2}$) & 7.7 & & \\
  $N_{\rm r}$  ($\times 10^{21}$cm$^{-2}$) & 9.7 & & \\
  $M_{\rm b}$  ($M_{\odot}$) & 10.7 &  &  \\
  $M_{\rm r}$  ($M_{\odot}$) & 10.7 &  &  \\
  $M_{\rm out}$ ($M_{\odot}$) & 21.4 &  &  \\
  $P_{\rm out}$ (M$_{\odot}$ km s$^{-1}$) & 565 & 1038 & 3256 \\
  $E$  ($\times 10^{47}$ ergs) & 3 & 10.1 & 99 \\
  $t_{\rm dyn}$  ($\times 10^{3}$ yr) & 11.8 & 7.65 & 2.08 \\
  $\dot{M}_{\rm out}$  ($\times 10^{-3}M_{\odot}$ yr$^{-1}$) & 1.9 & 2.8 & 10 \\
  $\dot{P}_{\rm out}$  ($\times 10^{-2}M_{\odot}$\kms\ yr$^{-1}$) & 5 & 14 & 157 \\
  $L_{\rm m}$ ($L_{\odot}$) & 210 & 1089 & 39470 \\
\hline
\end{tabular}
\end{center} 
\end{table}

In Table~\ref{outflow_tab}, the characteristics of the flow are given.
Col.~1 lists the computed parameters which are: the H$_2$ column density of the 
blue- and red-lobe, $N_{\rm b}$ and $N_{\rm r}$; the mass of the blue-
and red-lobe, $M_{\rm b}$ and $M_{\rm r}$, and the total mass, $M_{\rm out}$;
the momentum, $P_{\rm out}$; the energy, $E$; the dynamical timescale,
$t_{\rm dyn}$; the mass entrainment rate, $\dot{M}_{\rm out}$; the
mechanical force, $\dot{P}_{\rm out}$; the mechanical luminosity,
$L_{\rm m}$.

The H$_2$ column densities in both the blue and red lobe, $N_{\rm b}$ and
$N_{\rm r}$, have been derived from the 
\CO\ (2--1) line integrated emission in the wings according to the 
standard relations (see e.g.~Rohlfs \& Wilson~\citeyear{rew}):
\begin{equation}
N_{\rm b,r}=\frac{\rm H_2}{\rm CO} \frac{N_{\rm J(b,r)}}{g_{\rm J}}Q \exp({E_{\rm J}/T_{\rm ex}}) \;,
\label{coldenstot}
\end{equation}
\begin{equation}
N_{\rm J(b,r)}=\frac{3 k}{8 \pi \nu \mu^2}\frac{2J + 1}{J}\frac{\tau}{1-\exp(-\tau)}\int_{\rm b,r} {T_{\rm MB}{\rm d}V}\;.
\label{coldenspar}
\end{equation}

In Eq.~(\ref{coldenstot}),$\frac{\rm H_2}{\rm CO}$ is the inverse 
relative abundance of CO (assumed to be 10$^4$, Frerking et 
al.~\citeyear{frerking}), $N_{\rm J}$, $g_{\rm J}$ and $E_{\rm J}$ are
the column density, the statistical weight and the energy (in K) of the 
upper level (i.e.~$J=2$), respectively,
$Q$ is the partition function and $T_{\rm ex}$ the excitation temperature.
For this latter, we have assumed the dust temperature, i.e. 30 K, assuming
coupling between gas and dust.
In Eq.~(\ref{coldenspar}), $k$ is the Boltzmann constant, $\mu$ the 
dipole moment (0.11 Debye for CO), $\nu$ the line rest frequency,
$\tau$ the optical depth and
$\int{T_{\rm MB}{\rm d}V}$ is the integrated intensity in the line
wings. The values listed in Table~\ref{outflow_tab} have been
derived assuming optically thin emission, i.e. 
$\frac{\tau}{1-\exp(-\tau)}\sim 1$.
By comparing the \CO\ and \CI\ emission in the wings, we
have found that this assumption is correct for our data.

All the other parameters given in Table~\ref{outflow_tab} have been 
derived according to Eqs.~(2) - (8) of Leurini et 
al.~(\citeyear{leurini}), which take into account for correction for the flow 
inclination angle, $\theta$, with respect to the line-of-sight. The 
parameters are given for no correction for the 
inclination angle, for $\theta=57 \degr$ (the mean inclination
angle given by Cabrit \& Bertout~\citeyear{ceb}), and for 
an extreme value of $\theta=80 \degr$. Since the outflow orientation
looks close to the plane of the sky, we will consider the values computed
for $\theta=57 \degr$  and $\theta=80 \degr$ as lower and upper limits,
respectively. In this range, the flow parameters (outflow mass, momentum, 
energy, mechanical force and mechanical luminosity) are
consistent with the results obtained for outflows associated with 
high-mass YSOs observed at high-angular resolution (e.g.~Beuther et 
al.~\citeyear{beuther04}, ~\citeyear{beuther06}). 

We will better discuss 
the nature of the exciting source of the CO outflow in Sect.~\ref{nat_cont}.

\section{Discussion}
\label{discu}

\subsection{Nature of the continuum sources}
\label{nat_cont}

The maps shown in Sect.~\ref{cont} and the parameters derived in
Sect.~\ref{par_dust} allow us to discuss the nature of the millimeter
continuum cores C1-b, C1-a and C2.

C1-b probably hosts a very young early-B ZAMS star. This is indicated by 
several pieces of evidence:
First, Molinari et al.~(\citeyear{mol02}) have detected towards \i\
a cm-continuum source whose position roughly
corresponds to that of C1-b, and the Lyman continuum derived from
their observations suggests that the embedded object is a newly
formed B2--B3 ZAMS star. Second, the source is embedded in a
hot-core, as demonstrated by the kinetic temperature derived
from CH$_3$CN (12--11) in Sect.~\ref{ch3cn}, and by the presence of 
other high-excitation lines (see Fig.~\ref{figura_coetal}), 
and it lies where the main peak of the mid-infrared 24 $\mu$m continuum emission
is detected (see Fig.~\ref{cont_spi_fig}). Third, it is close to the center of the outflow
lobes detected in \CO\ (2--1) (Fig.~\ref{cooutflow}), even though
a significant contribution to the observed blue- and
red-shifted emission can arise from C1-a and, less likely, from another source 
not seen in the continuum image, located at the position of the
western emission peak of \H\ (see Fig.~\ref{cooutflow}). As discussed in Sect.~\ref{par_dust},
even though the masses derived form the millimeter continuum are
just 5.2 and 0.6 \solm , these measurements represent the mass of the circumstellar gas
only, and thus they do not contrast with the previous findings.

The source embedded in C1-a is 
probably a very young intermediate-mass protostar. This is
indicated by: the core location, at the centre of the \CO\ outflow lobes,
which suggests that the object embedded in the core is one of the main sources
driving the powerful outflow; the core gas mass listed in Table~\ref{cont_tab}; the
non-detection of the core in the centimeter continuum 
(Molinari et al.~\citeyear{mol02}), indicating an evolutionary stage
prior to the formation of an HII region. Based on these findings, we believe
that the source embedded in C1-a is an intermediate-mass protostar
still in the main accretion phase (i.e. a Class 0 object).

The nature of the source embedded in C2 is unclear. The core mass
(see Table~\ref{cont_tab}) and the non-detection in the 24 $\mu$m
image suggest that it can be an intermediate-mass embedded protostar. 
The source is not associated with a detectable outflow, which however could be
an indication that the embedded source is extremely young. Interestingly, one can notice
from Fig.~\ref{map_thin} that C2 is in the middle of a filamentary structure
extended in the N-S direction which also contains the deuterated cores N and S,
and four of the five peaks of the \H\ (1--0) integrated emission. This may
indicate that this portion of the source host the youngest low- and intermediate-mass 
objects of the proto-cluster, and it is consistent with theoretical models
that predict that filamentary structures are common morphologies of molecular
clouds in several classes of dynamical formation models (e.g. van Loo
\citeyear{vanloo}).

\subsubsection{A proto-Trapezium system?}

From the positions given in Table~\ref{cont_tab}, we derive that
the projected linear separation among the three continuum sources
goes from $\sim 0.014$ pc between C1-a and C1-b, to $\sim 0.07$ pc 
between C1-a and C2, i.e. from $\sim 2800$ to 14000 A.U., at the given distance.
Looking at the 284 and 225 GHz images in Fig.~\ref{cont_fig}, one could also argue the possible 
presence of a fourth continuum unresolved source to the NW of C1-a. 
This is suggested by the elongated shape of C1-a in this direction, while 
the synthesised beam of the SMA at 284 GHz is elongated in NE - SW direction.
Assuming spherical symmetry, the stellar density
in this region is thus $\sim 2.3 \times 10^{4}$ proto-stars pc$^{-3}$, 
while it is $\sim 1.7. \times 10^{4}$ proto-stars pc$^{-3}$ assuming the three 
resolved sources only. These values are consistent with the stellar density measured in the 
Trapezium cluster in Orion, estimated to be
of the order of 10$^4$ stars pc$^{-3}$, and derived assuming the lower-mass stars
in the cluster as well. This suggests that the stellar density of $\sim 1.7 \times 10^{4}$ proto-stars pc$^{-3}$
is a lower limit, and that perhaps the three young intermediate-/high-mass (proto-)stars embedded 
into C1-a, C1-b and C2 (plus possibly the unresolved one to the NW of C1-a)
represents a Trapezium-like system in the making.

Similar multiple systems of forming intermediate-/high-mass stars have been discovered 
recently through high-angular resolution observations (see e.g. Beuther et al.~\citeyear{beuther07}; 
Rodon et al.~\citeyear{rodon}), although these systems are denser than the one in
\i\ (more than $10^5$ stars pc$^{-3}$). Neverthless, these stellar densities are much lower 
than the value of $10^8$ stars pc$^{-3}$ needed for the formation of high-mass stars 
through merging of several lower-mass stars at the center of rich clusters 
(see e.g.~Bonnell et al.~\citeyear{bonnell98}), but comparable to the value
sufficient to produce binary induced mergers (Bonnell \& Bate~\citeyear{beb}).

\subsection{Nature of the deuterated cores}
\label{nat_det}

In paper I, we discussed the nature of the deuterated cores
based on the dust continuum emission and the deuterium
fractionation derived from the column density ratio $N$(\D )/$N$(\H).
We derived the mass of the cores following
two approaches: from the virial theorem, and assuming
an average abundance of \H . For completeness,
in Col.~6 of Table~\ref{tab_dfrac} we list the main results
obtained from the second approach in paper I. 
These values, together with a \Dfrac\ $\simeq 0.1$ (Col.~4 of
Table~\ref{tab_dfrac}), 
suggested that both cores are in the pre--stellar phase
but at present they are not massive.
On the other hand, as discussed in Sect.~3.3 of paper I, 
the average spectra of the \H\ (1--0) and \D\ (3--2) lines
in both N and S show lines broader than those typically 
observed in low-mass pre--stellar cores. 

We proposed two possible scenarios for the nature of the two \D\
condensations:  (i) they are quiescent low-mass 
starless cores, and therefore they are going to form low-mass stars
(in particular, N is likely to harbor several unresolved cores); 
(ii) they are condensations still in dynamic evolution, which can be
changing their mass and shape, undergoing either fragmentation
or accretion from the parental cloud. In particular, the second scenario is 
suggested by the predictions of the 'competitive accretion models',
which predict formation of massive stars in a clustered
environment through dynamical interaction among several 'seeds' of forming
(proto-)stars (see e.g.~Zinnecker \& Yorke~2007 for a review). 

To better understand the nature of the two condensations, a 
detailed analysis of the gas kinematics is certainly very helpful. 
For this reason, here we investigate in more detail the gas kinematics in 
N and S by using as diagnostics the peak velocity and  widths 
of the \H\ (1--0)  and \D\ (3--2) lines, following the same approach described
in Sect.~\ref{par_kin}.

\subsubsection{Condensation N}

 In the two left panels of Fig.~\ref{fig_kinN} we show the maps of $V_{\rm LSR}$
and $\Delta V$, respectively, obtained for N from the \H\ (1--0) line.
The same plots derived from \D\ (3--2) are shown in the two right panels.
One can see in the first place that the velocity field in this
core is quite complex. The line width of the \H\ (1--0) line are slighly larger
in the southern region of the core, while that of the \D\ (3--2) looks 
larger in the northern and, less prominently, in the south-eastern portion of the core.
For both species, the values range from $\sim 0.6$ to $\sim 1.4$ \kms ,
and are on average 3--4 times larger than those found in low-mass
pre--stellar cores (see e.g. Caselli et al.~2002a for the case of
L1544, Crapsi et al.~2005). However, for the \D\ (3--2) line, in the positions
where $\Delta V \sim 0.5$ \kms , these values can be influenced by the
poor spectral resolution of $\sim 0.5$ \kms\ (see Table~\ref{tab_mol}). 
Therefore, it is possible that linewidths smaller than
0.5 \kms\ are present but not observable with the adopted spectral resolution. 
The peak velocity of both \H\ and \D\ lines are between $\sim -17.7$
and $\sim -18.4$ \kms , i.e. close to the systemic
velocity, but the gas in the southern region of the core looks slightly red-shifted
with respect to the northern one. 
Also, from Fig.~\ref{cooutflow} one can notice that the red
lobe of the \CO\ outflow, likely driven by C1-b, arises from a region
spatially coincident with the southern portion of core N, and possibly
interacts with it: the gas in the red-shifted lobe
of the outflow probably flows behind the core, dragging on part
of the southern gas of N. In this way, one can
explain the observed red-shift of the \H\ and \D\ lines in the
southern portion of N. This could also be interpreted as being due to
the presence of unserolved low-mass cores at a different velocity, and 
the \H\ emission peaks shown in Fig.~\ref{map_thin} is consistent with 
this interpretation. In this case, the observed broad lines are just due to
the superposition of the different cores. However, this seems very
unlikely since the red-shifted gas in the southern portion of the
core is spatially coincident with the bulk of the red lobe of the
\CO\ outflow. So that we believe that the observed red-shifted emission
and broad lines in the southern portion of N are both due to the interaction with the ouflow.

\subsubsection{Condensation S} 

\begin{figure*}[t!]
 \begin{center}
 \resizebox{\textwidth}{!}{\includegraphics[angle=-90]{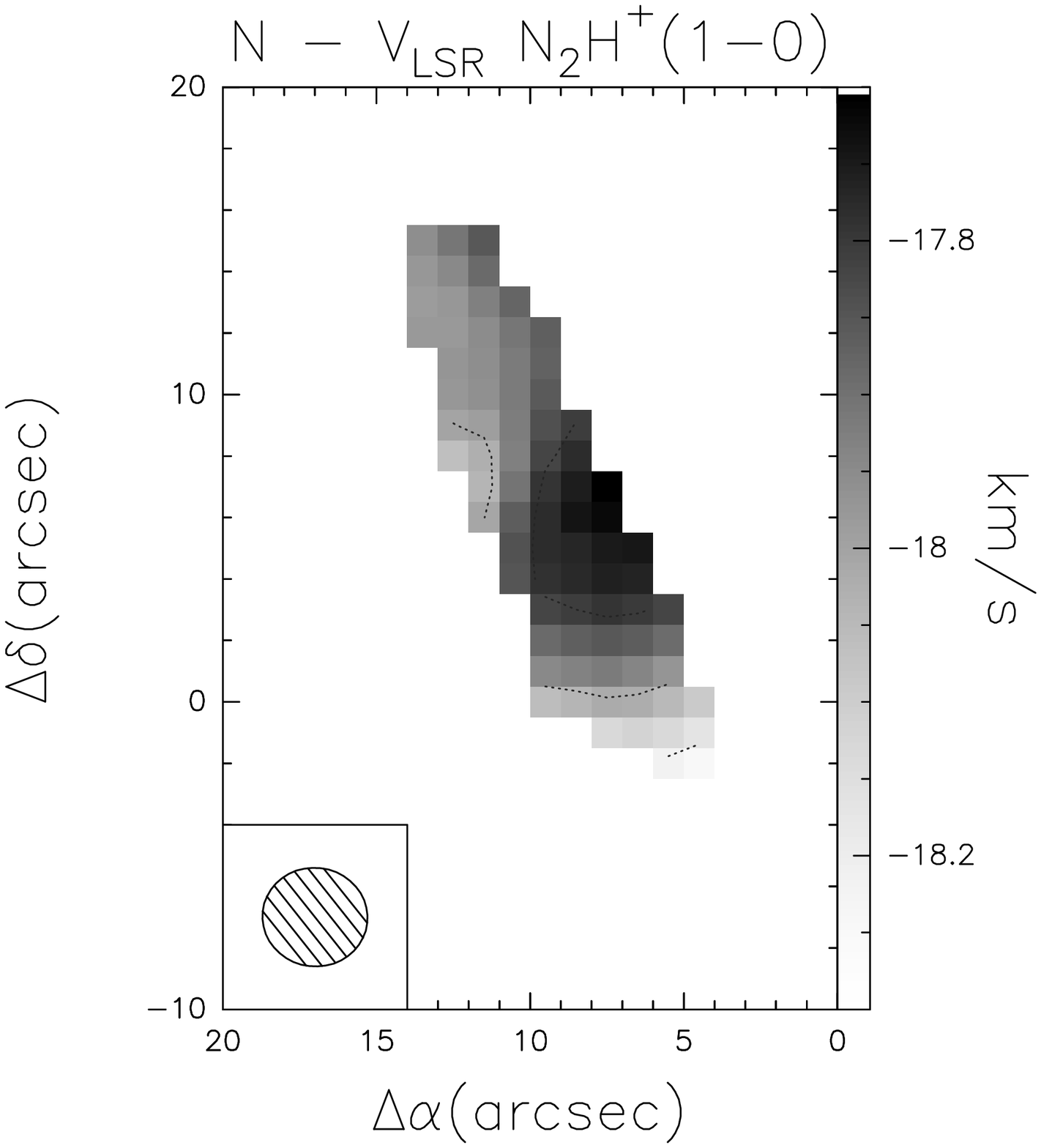}%
 			   \includegraphics[angle=-90]{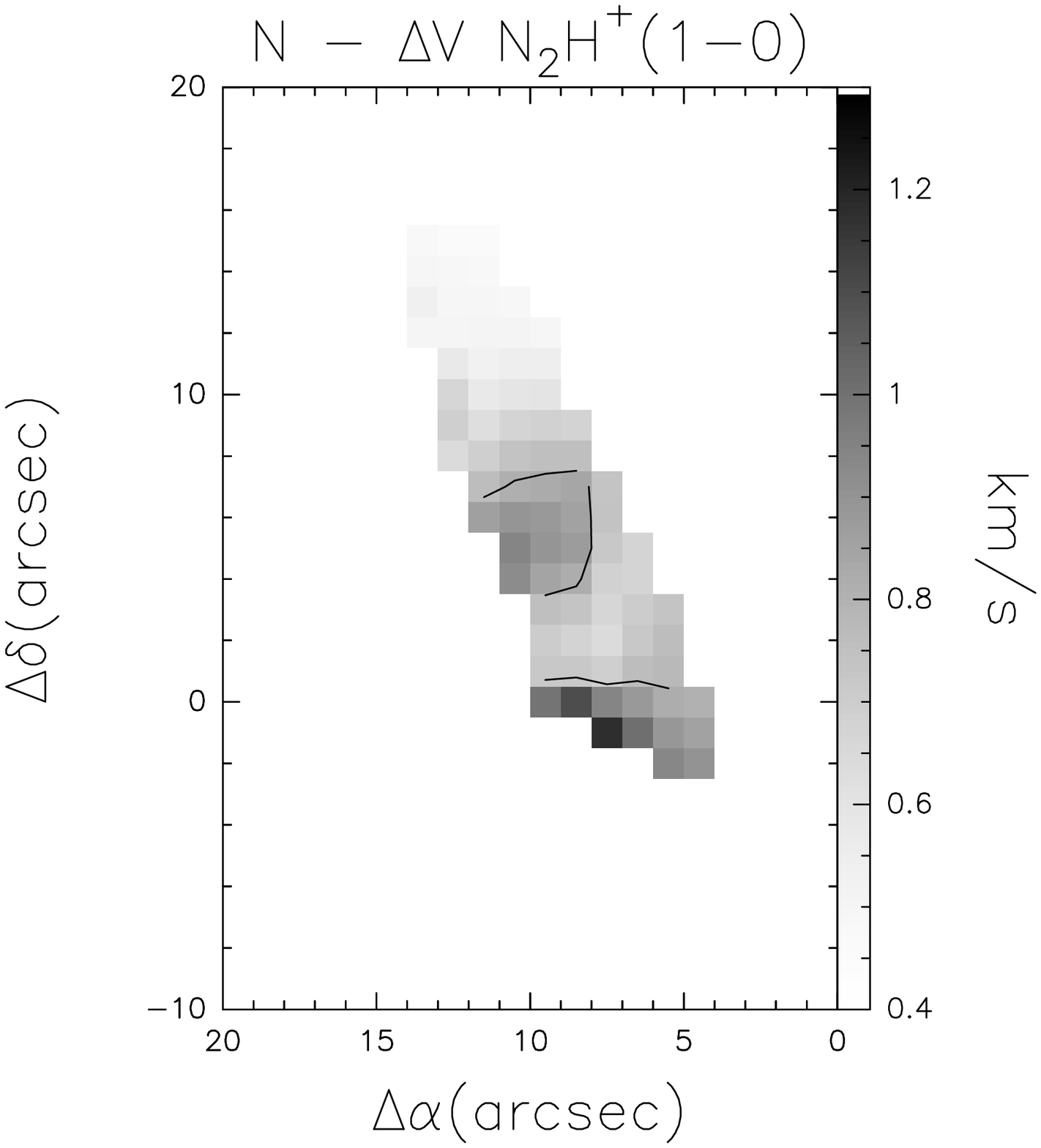}%
                           \includegraphics[angle=-90]{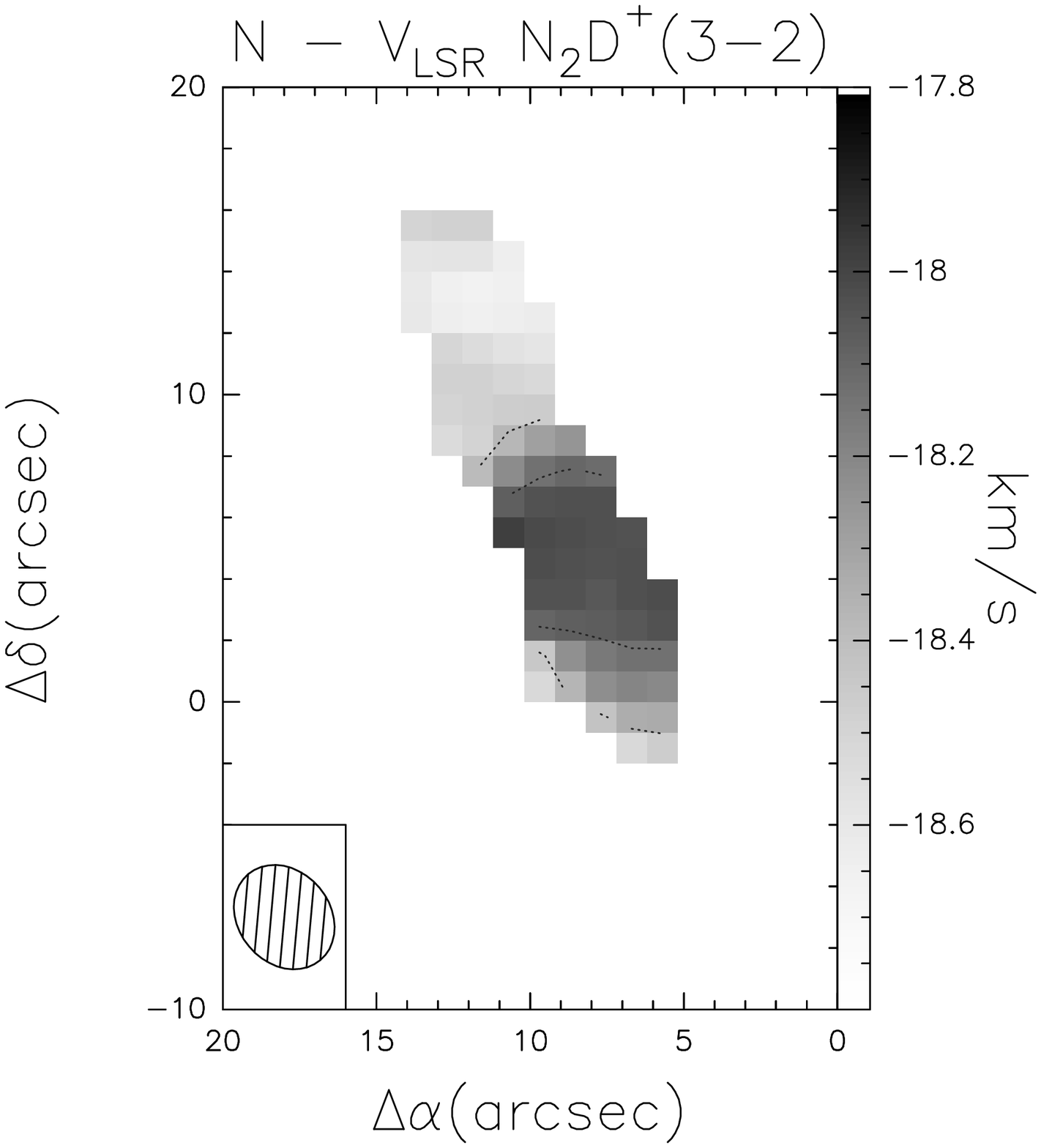}%
                           \includegraphics[angle=-90]{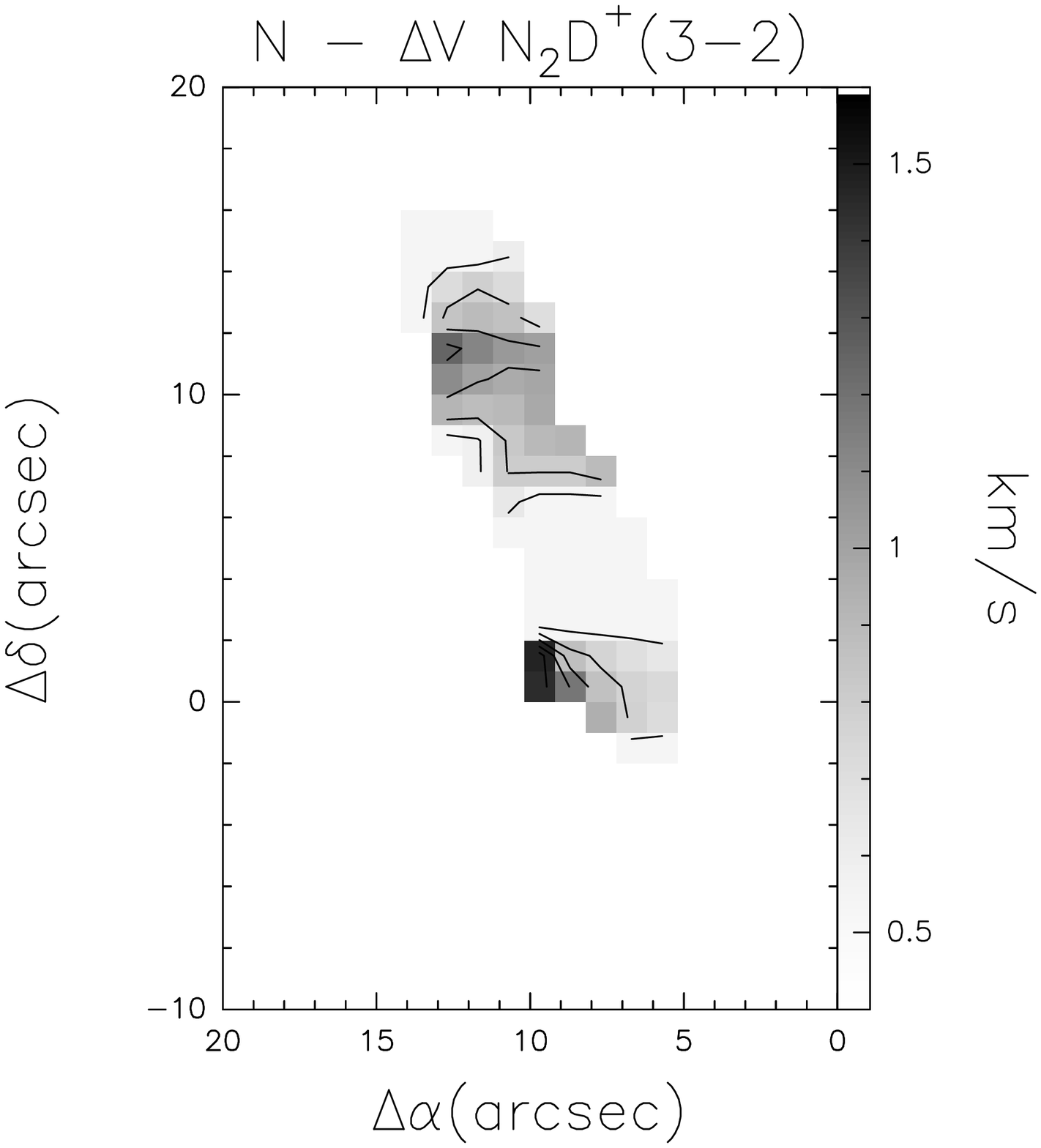}} 
 \end{center}
 \caption
 {\label{fig_kinN} Left panels: map of the peak velocity ($V_{\rm LSR}$) and
 line width ($\Delta V$) derived from the \H\ (1--0) line inside the 3$\sigma$ 
 level of the \D\ (3--2) emission of core N. 
 The vertical grey-scale on the
 right side of the two plots indicates the intensity, in \kms , of $V_{\rm LSR}$
and $\Delta V$, respectively. In the bottom left corner, the synthesised beam
of the channel maps are shown. Right panels: same as left panel
 for the \D\ (3--2) line.}
\end{figure*}

\begin{figure*}[t!]
 \begin{center}
 \resizebox{\textwidth}{!}{\includegraphics[angle=-90]{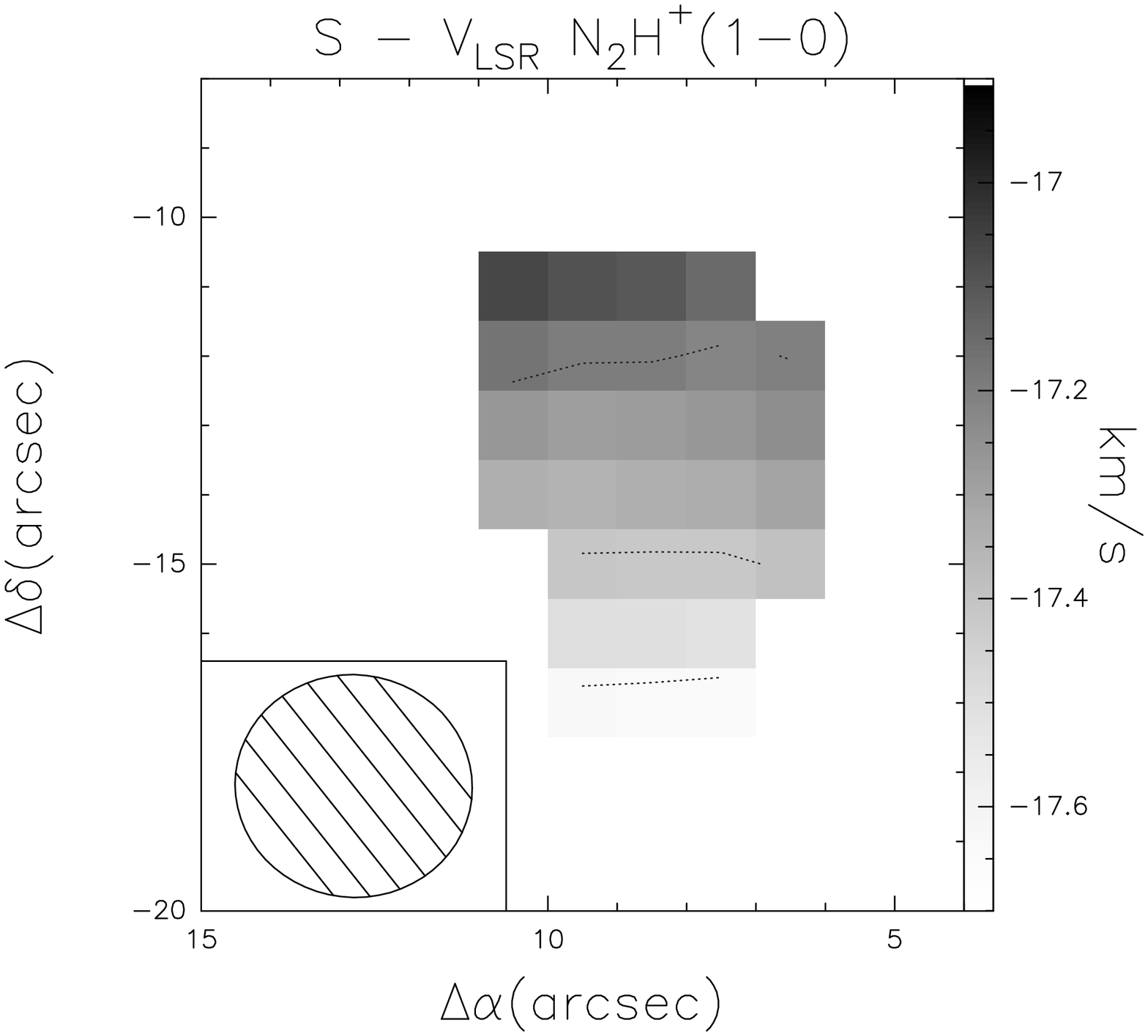}%
 			   \includegraphics[angle=-90]{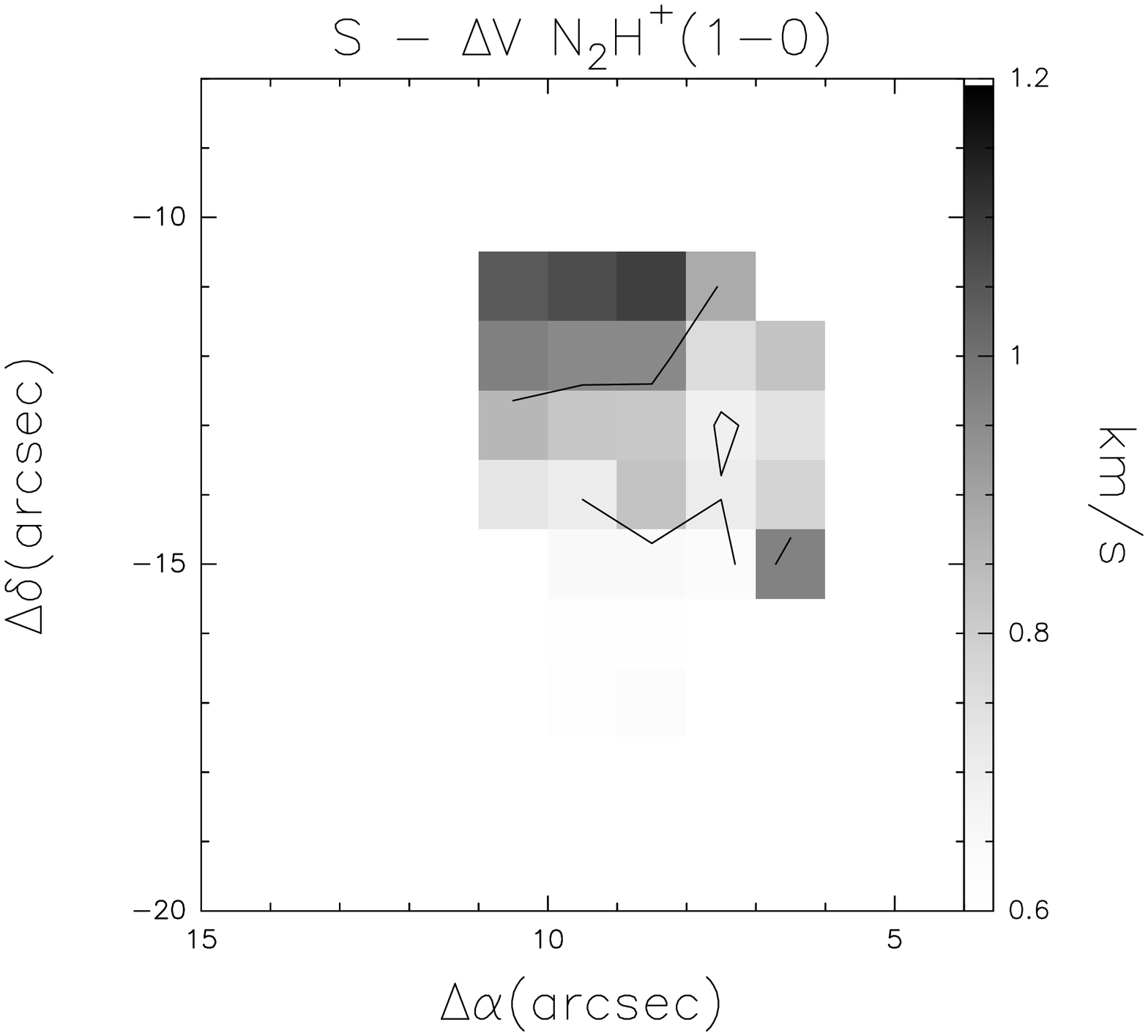}%
                           \includegraphics[angle=-90]{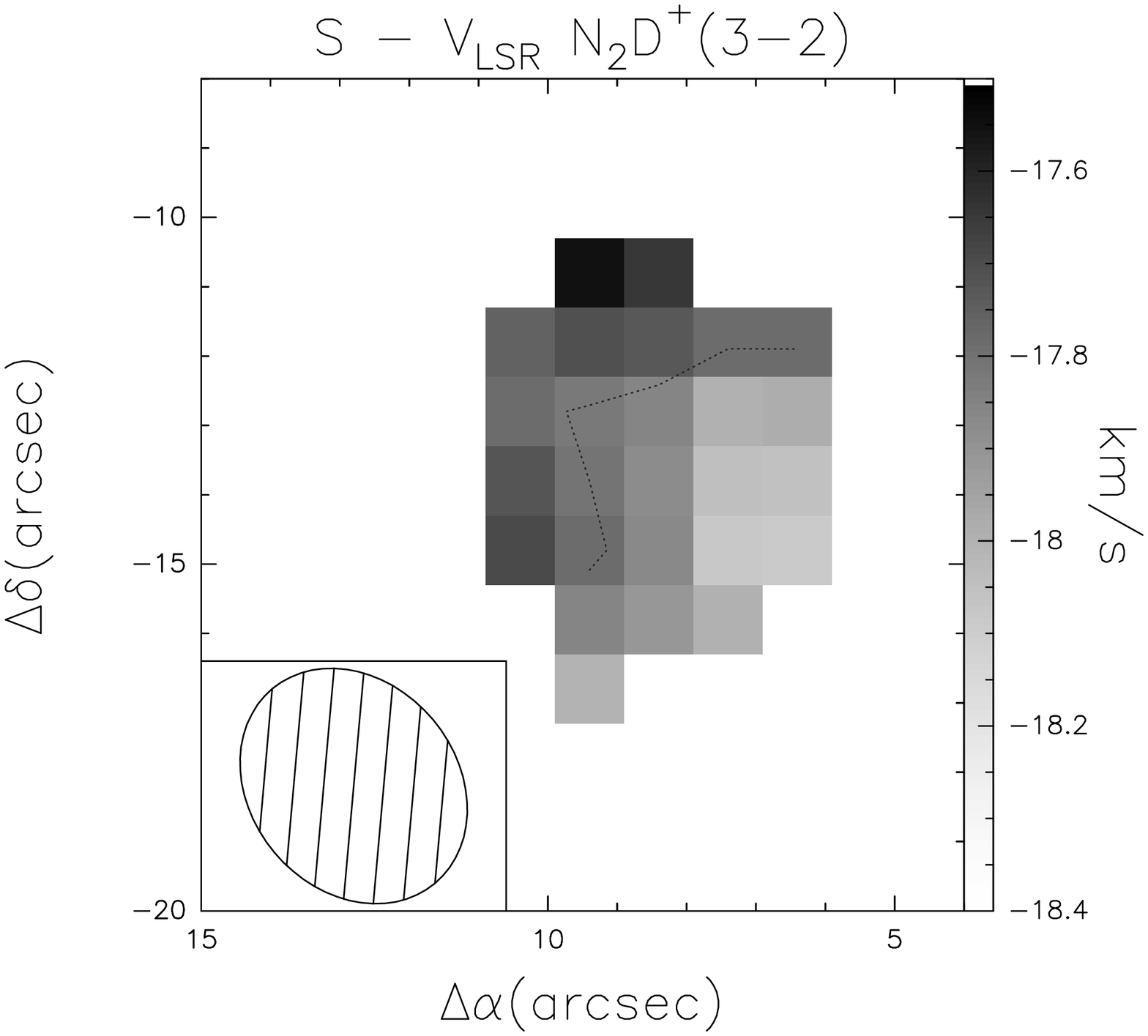}%
                           \includegraphics[angle=-90]{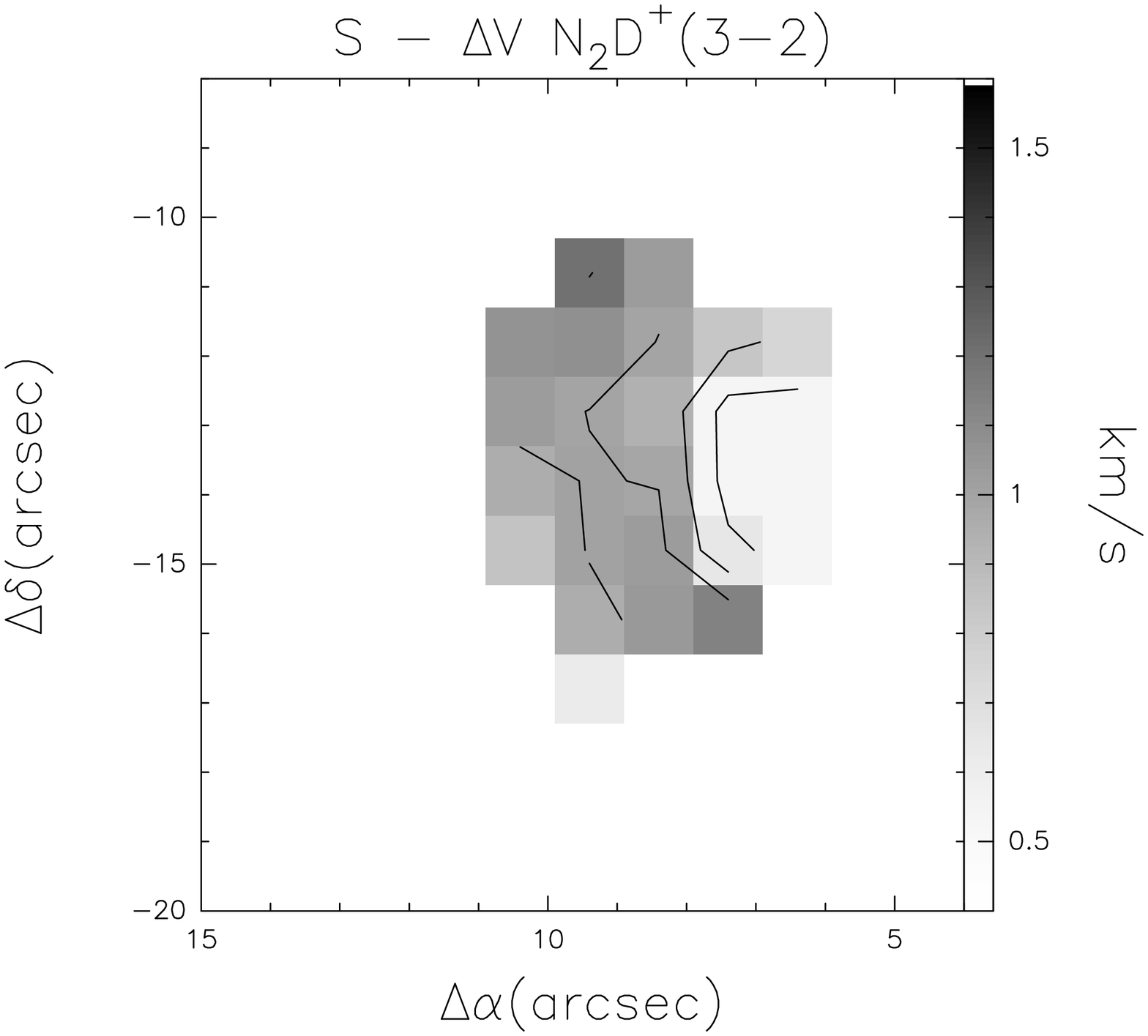}} 
 \end{center}
 \caption
 {\label{fig_kinS} Same as Fig.~\ref{fig_kinN} for core S.}
\end{figure*}

In Fig.~\ref{fig_kinS} we show the same maps as
in Fig.~\ref{fig_kinN} for core S. The peak velocity of
both the \H\ (1--0) and \D\ (3--2) lines look red-shifted towards 
the north of the core (which appears to be more shifted to
the north-east in \D ), suggestive of a rotation motion.
A disk-like structure has been observed in the low-mass pre--stellar
core L1544, and it is expected to be caused by the ambipolar
diffusion mechanism (Ciolek \& Basu~\citeyear{ceb}; Caselli et al.~2002a). 
However, L1544 looks 
also flatter than what is seen in core S, which shows a spherical shape.
This can be due to different orientation with respect to the line of sight
and to the worse angular resolution of our data, since Caselli
et al.~(2002a) obtained observations with linear resolution a factor of $\sim 3$
better than ours. Interestingly, $\Delta V$ of both \H\ and \D\ are
larger towards the north-eastern part of the core, so that the
red-shifted gas appears to be also more turbulent. As for N,
this is suggestive of an interaction of the core with
the red lobe of the \CO\ outflow, the southern edge of which
touches the upper side of core S (see the upper panel of Fig.~\ref{cooutflow}).

In summary, the shape and the mass of this core indicate
that it is a low-mass pre--stellar core, but the high 
turbulence and the red-shifted gas revealed in the northern part of the core
(probably triggered by the neighbouring \CO\ outflow)
can influence the evolution of the core itself. 
\vspace {0.5cm}
                            
\subsubsection{General conclusion on the nature of N and S}
                            
The main finding of the analysis performed above is that both
cores are characterised by a probable interaction with
the red lobe of the \CO\ outflow. This can trigger turbulence in the
cores themselves and can influence their evolution.
In theoretical models of clustered star formation, turbulence (which
in this case is generated by already formed protostars)
can create density modifications across the cloud,
originating several dense and cold 'seeds', which subsequently can
accrete backgound gas that was initially not associated with the 
'accretion domain' of the seed itself
(see e.g. Bonnell et al.~\citeyear{bonnell}, and McKee \& Tan
~\citeyear{mckee2}). 
In this scenario, the two condensations can become more massive
and form pre--stellar cores of higher mass. Alternatively, the
interaction with the other cluster members, in particular with
the powerful \CO\ outflow, could also cause
the fragmentation of the condensations. 

\subsection{Comparison with other star forming regions}
\label{comparison}

\subsubsection{Intermediate-/high-mass}
\label{highmass}

The analysis made in Sects.~\ref{nat_det} and \ref{nat_cont} indicates that
in \i\ there are objects with masses between $\sim 2$ and $\sim 20$ \solm ,
and in different evolutionary stages: pre--stellar core candidates, intermediate-mass
class 0 protostars, and a newly formed early-B ZAMS star. In recent years, many
star forming regions harboring intermediate- and high-mass young stellar objecs
have been investigated at high-angular resolution, and several of these regions
show a complex structure as seen in \i . However, only in very few of them,
evidence of the presence of pre--stellar core candidates has been found.

Among the studies performed so far, \i\ appears to have the strongest similarities 
with the high-mass protocluster associated with IRAS 20343+4129.
Palau et al.~(\citeyear{palau}) have observed this cluster with the SMA in the
sub-mm continuum and in \CO\ (2--1). Since the synthesised beam and the
source distance are $\sim 3$ \asec\ and 1.4~kpc, respectively, their
observations have linear resolution comparable to our ones. They have found
two intermediate- to high-mass young stellar objects, one of which is driving a molecular
outflow, and the other one is driving an expanding cavity. Interestingly, the gas 
at the edge of the cavity seems to be compressed into
several dense cores, detected in the millimeter continuum only.
Palau et al.~(\citeyear{palau}) propose that these are potential starless 
cores representing a future generation of stars. In this scenario,
the formation of new stars in the cluster is caused by the interaction
between the highest mass objects (which are expected
to form first at cluster center), and the residual circumstellar (circum-cluster)
material as in \i . 

Leurini et al.~(\citeyear{leurini2}) have investigated
at high-angular resolution the high-mass star forming
region IRAS 05358+3543, and found 4 massive millimeter
cores. One of these seems to be 
in the pre--stellar phase, being very cold ($T\leq 20$ K) and with a large
reservoir of material ($M\sim 19$ \solm ). However, the core does not show
any interaction with the other cluster members, and the
authors suggest that it could be also a very embedded massive protostar.

\subsubsection{Low-mass}
\label{lowmass}

Several well-studied low-mass star forming regions show
filamentary structures similar to that seen in \i , and clustered star formation. 

In the $\rho$-Ophiuchi molecular cloud complex, the
large majority of low-mass millimeter cores associated with both pre-- and 
proto--stellar objects are found to be distributed into filaments (see
e.g. Nutter et al.~\citeyear{nutter}, Andr\'e et al.~\citeyear{andre}).
In particular, the structure of one of these filaments, Oph A, closely
resembles that of core N. Andr\'e et al.~(\citeyear{andre})
have used \H\ (1--0) to perform a detailed study of the kinematics in 
Oph A, and their results are different from ours: first,
they derive an average line width (thermal +
non-thermal component) of $\sim 0.4$ \kms , which is yet more than
2 times smaller than that measured in our deuterated cores, even
though in the \D\ lines we are limited by a spectral resolution of 0.5 \kms ,
and we are at the limit of our resolution towards some parts of the
deuterated cores .
Second, they do not find any evidence of interactions between
the different cores, so that there is not feedback from protostellar
activity. Similar morphologies have been obtained
observing \H\ (1--0) towards the Perseus molecular cloud by
Kirk et al.~(\citeyear{kirk}), in which there seem to be a minimal
contribution of turbulence in the observed line widths.

These results clearly indicate that pre--stellar cores in low-mass star
forming regions are commonly found in
filaments similarly to what we have found in \i , but they are much less turbulent
and less interacting with the other cluster members than those
studied in this work.
Therefore, as suggested in Sect.~\ref{nat_det}, in the condensations
associated with \i\ the nearby intermediate- and high-mass
young objects likely play an important role in triggering
turbulence and, eventually, in causing either the growth or 
the dispersion of the gas in these cores. 

\section{Summary and conclusions}
\label{conc}

We have presented a full report of the observations partially analysed by Fontani
et al.~(\citeyear{fonta08} - paper I) towards the intermediate-/high-mass star forming region 
IRAS 05345+3157. The observations have been performed with the PdBI and SMA 
in the following molecular transitions: \H\ (1--0) with PdBI, \H\ (3--2) and 
\D\ (3--2), both with the SMA. At the frequencies of these lines, we have
simultaneously observed the continuum emission, as well as other rotational
transitions as \CO\ (2--1), \CI\ (2--1), \CII\ (2--1) and few other high excitation
lines of less abundant species. In paper I, the main finding presented was the
detection of two molecular condensations (called N and S) showing high values of deuterium
fractionation ($\sim 0.1$), derived from the column density ratio $N$(\D )/$N$(\H ).

In this work, the following main results have been obtained:
\begin{itemize}
\item The continuum emission at 96 and 225 GHz reveals two main cores, C1 and C2,
and C1 is resolved into two components in the image at 284 GHz, 
called C1-a and C1-b. C1 is coincident with a prominent 
mid-infrared emission detetcted in the Spitzer MIPS images at 24 and 
70 $\mu$m, which is only barely detected towards C2. 
\item The integrated intensity of the optically thin component of the \H\ (1--0) line
shows an extended distribution with 5 emission peaks.
Two of them fall inside condensation N, another one at the
edge of condensation S, while none of them
overlaps with the continuum sources.
\item The integrated emission in the \CO\ (2--1) line wings reveals the
presence of a powerful bipolar outflow oriented roughly in the WE direction.
From the outflow geometry, the sources C1-a and C1-b are the best candidates
for powering the outflow.
Assuming a unique source driving the outflow, its parameters are consistent
with an engine which is a high-mass young stellar object.  
\item The \H\ (1--0) line widths are between 1 and 2
\kms\ in a region that spatially corresponds to that where
the continuum cores are located, while they are significantly
smaller (between 0.5 and 1.5 \kms ) in the deuterated cores. 
\item Based on previous observations and on the results
presented in this work, we can conclude that C1-b very likely harbors
a newly formed early-B star embedded inside a hot-core, and C1-a 
harbors an intermediate-mass class 0 protostar. The nature of C2 is unclear
but it could be a very embedded intermediate-mass protostar. If we consider the 
system of these three continuum sources and roughly assume a spherical
symmetry, we deduce a star density of $\sim 10^{4}$ stars pc $^{-3}$,
consistent with the stellar density of the Trapezium cluster and much lower
than the value required to form massive stars through collisions of
low-mass ones.
\item The nature of the deuterated cores has been discussed using mainly
the information on the gas kinematics that one can derive from the lines of 
\H\ and \D . Core S is likely a 
single low-mass pre--stellar core, and the velocity of both the \H\ and \D\ lines
indicate a rotation motion roughly in the N-S direction. This motion
and the high turbulence
measured at the northern edge of the core is likely due to dynamical
interaction with the red lobe of the \CO\ powerful outflow.
The nature of core N is less clear. Two scenarios are possible:
it can be a filament of unresolved low-mass pre--stellar cores, similarly to
Oph A, and in this case the high values of the \H\ and \D\ line widths
are just due to the superposition of the several unresolved cores.
Alternatively, it can be a single object still in dynamic evolution, which actively
interact with the red lobe of the \CO\ outflow. In this case, the triggered
trubulence can cause either the growth or the fragmentation of the core itself.
\end{itemize}

From these observations, it is not at all clear how the interaction with the other
cluster members can make the condensations evolve towards the formation
of low- or high-mass objects. Only higher resolution kinematic studies of the large 
scale gas around these condensations, and the comparison of these results with
hydrodynamic evolution models, will help to understand the core evolution and
provide important constraints on the dynamical evolution of intermediate-/high-mass
star forming regions.

\vspace{1cm}

{\it Acknowledgments.} First of all, many
thanks to the anonymous referee for his/her useful comments and suggestions.
It is a pleasure to thank the staff of the 
Smithsonian Astrophysical Observatory for the SMA observations. We 
also thank the IRAM staff and Riccardo Cesaroni for their help in the calibration 
of the PdBI data. FF is deeply grateful to Marc Audard, Carla Baldovin and 
Andres Carmona for useful discussion, and to Keping Qiu for fitting the 
CH$_3$CN spectrum. PC thanks Jonathan Tan for useful discussions
on protocluster dynamics. FF acknowledges support by Swiss
National Science Foundation grant (PP002 -- 110504). TB acknowledges support
from US National Science Foundation grant (0708158).
{}

\end{document}